\documentclass[10pt]{article}
\usepackage{graphicx}
\usepackage{amsmath}
\usepackage{amssymb}
\usepackage{caption2}
\begin{document}
\bibliographystyle{prsty}
\begin{center}
{\large {\bf \sc{  Analysis of the $B \to K^*_2(1430),
a_2(1320) , f_2(1270)$ form-factors with light-cone QCD sum rules  }}} \\[2mm]
Zhi-Gang Wang \footnote{ E-mail,wangzgyiti@yahoo.com.cn.  }    \\
 Department of Physics, North China Electric Power University, Baoding 071003, P. R. China
\end{center}

\begin{abstract}
In this article, we study the $B \to K^*_2(1430)$, $a_2(1320)$,
$f_2(1270)$ form-factors  with the  light-cone QCD sum rules, where
the $B$-meson light-cone distribution amplitudes are used. In
calculations, we observe that the line-shapes of the $B$-meson
light-cone distribution amplitude $\phi_+(\omega)$ have significant
impacts on the values of the form-factors, and expect to obtain
severe constraints on the parameters of the $B$-meson light-cone
distribution amplitudes from the experimental data in the future.
\end{abstract}



\section{Introduction}

The semi-leptonic and radiative
$B$-decays to the light tensor mesons   play an  important role in testing  the standard
model and searching for new physics. At the quark level,  the transitions $b\to s(d)\gamma$ and $b \to s(d)\ell^+\ell^-$ occur through the
flavor changing neutral currents, which are forbidden at the
tree-level in the standard model, provide fertile ground for testing
the standard model at the loop-level as well as obtaining useful
information about new physics effects. The $B\to T$ form-factors
enter the semi-leptonic decays  and radiative  decays, and
serve as the basic parameters,  their  values with improved
precision can reduce the theoretical uncertainties. Experimentally,
the radiative decays $B^0 \to K^{*0}_2(1430) \gamma$ and $B^+ \to
K^{*+}_2(1430) \gamma$ have been observed in the past years
\cite{BK2-1,BK2-2}. Other radiative decays and semi-leptonic decays
are expected to be observed  at the KEK-B and the LHCb in the
future. The tensor mesons cannot be produced from the
local $V\pm A$ currents of the standard model, the $B\to T$ form-factors also play an important role in studying the
two-body hadronic decays $B\to TM$ (with $M=P,V,A$), the calculations
based on the naive factorization approach and QCD-improved
factorization approach cannot account for the experimental data
satisfactorily \cite{ChengYang}, the non-factorizable contributions maybe large enough.
 Comparing with the two-body hadronic $B$-decays, the semi-leptonic and radiative
$B$-decays suffer from much less theoretical uncertainties involving   the hadronic matrix elements.

 The $B\to T$ form-factors concern  the nonperturbative sector of QCD and
have been calculated  in the ISGW model \cite{ISGW} and its improved
version (the ISGW2 model) \cite{ISGW2}, the covariant light-front
quark model \cite{CCH}, the light-cone sum rules approach (where the
light-cone distribution amplitudes of the tensor  mesons are used)
\cite{kcy}, the large energy effective theory
\cite{Charles,Ebert,Datta} and the perturbative QCD approach
\cite{Wei}, etc. Another calculation based on the light-cone QCD sum
rules is useful.

In Refs.\cite{Khodjamirian05,KhodjamirianB07},  Khodjamirian et al
obtain new sum rules (the so-called  $B$-meson light-cone QCD sum
rules) from the correlation functions expanded near the light-cone
in terms of the $B$-meson distribution amplitudes for the $B\to
\pi,K ,  \rho,K^*$ form-factors.   Furthermore, they suggest QCD sum
rules motivated models for the three-particle $B$-meson light-cone
distribution amplitudes, which satisfy the relations between the
two-particle and three-particle $B$-meson light-cone distribution
amplitudes derived from the QCD equations of motion and heavy-quark
symmetry \cite{Qiao2001}. The new QCD sum rules have been applied
successfully  to calculate the $B\to a_1(1260) \, , \, D\, , \, D^*$
form-factors, etc \cite{WangPLB,KhodjamirianEPJC}. In this article,
we use the $B$-meson light-cone QCD sum rules to study the $B \to
K_2^*(1430)$, $a_2(1320)$, $f_2(1270)$ form-factors.

 The $B$-meson light-cone distribution
amplitudes  play an important role in the exclusive $B$-decays, the
inverse moment of the two-particle light-cone distribution amplitude
$\phi_+(\omega)$  enters many factorization formulas
\cite{BBNS,Form4}.  The two-particle $B$-meson light-cone
distribution amplitudes have been studied  with the QCD sum rules
and renormalization group equation
\cite{Khodjamirian05,Grozin1997,Qiao2003,Lange2003,Braun2004,Grozin05,Lee05,Offen09}.
Although the QCD sum rules cannot be used for a direct calculation
of the distribution amplitudes, it can provide constraints which
have to be implemented within the QCD motivated models
 \cite{Braun2004}. Comparing with the light pseudoscalar mesons and
vector  mesons,  the $B$-meson light-cone distribution amplitudes
have  received relatively little attention. Our knowledge about the
nonperturbative parameters which determine those light-cone
distribution amplitudes is limited and an additional application (or
estimation) based on the light-cone QCD sum rules  is useful.

On the other hand, if those form-factors are extracted from the
experimental data on the semi-leptonic decays and radiative decays
at the KEK-B and LHCb in the future, we get  feedback, and obtain
severe constraints on the input parameters of the $B$-meson
light-cone distribution amplitudes. At the LHCb, the $b\bar{b}$
pairs will be copiously produced with the cross section about $500
\,\mu b$.

The article is arranged as: in Sect.2, we derive the $B \to
K_2^*(1430)$, $a_2(1320)$, $f_2(1270)$  form-factors      with the
light-cone QCD sum rules; in Sect.3, the numerical result and
discussion; and Sect.4 is reserved for our conclusion.

\section{$B \to K_2^*(1430)$, $a_2(1320)$, $f_2(1270)$  form-factors  with
light-cone QCD sum rules}

In the following, we write down the definitions for the $B\to
K^*_2(1430)$  form-factors $V(q^2)$, $A_1(q^2)$, $A_2(q^2)$,
$A_3(q^2)$, $A_0(q^2)$, $T_1(q^2)$, $T_2(q^2)$, and $T_3(q^2)$
\cite{Wei,T-Difin-1,T-Difin-2},
\begin{eqnarray}
\langle K^*_2(p)|\bar{s}(0)\gamma^\mu b(0)|B(P)\rangle&=&-
\frac{2}{m_B+m_{K^*_2}}\epsilon^{\mu\lambda\tau\rho}\epsilon_\lambda^*
P_\tau p_\rho V(q^2)\, ,\nonumber\\
 \langle K^*_2(p)|\bar{s}(0)\gamma_\mu\gamma_5 b(0)|B(P)\rangle&=&i\left\{
\left(m_B+m_{K^*_2}\right)\epsilon_\mu^*A_1(q^2)-\frac{\epsilon^*
\cdot
P}{m_B+m_{K^*_2}}(P+p)_\mu A_2(q^2) \right. \nonumber\\
&&\left.-2m_{K^*_2}\frac{\epsilon^* \cdot
P}{q^2}q_\mu\left[A_3(q^2)-A_0(q^2)\right]
\right\} \, ,\nonumber\\
\langle K^*_2(p)|\bar{s}(0)\sigma^{\mu\nu}q_\nu
b(0)|B(P)\rangle&=&2i \epsilon^{\mu\lambda\rho\tau} P_\lambda p_\rho
\epsilon_\tau^*T_1(q^2) \, ,\nonumber\\
\langle K^*_2(p)|\bar{s}(0)\sigma_{\mu\nu}\gamma_5q^\nu
b(0)|B(P)\rangle&=&T_2(q^2)\left[
\left(m^2_B-m_{K^*_2}^2\right)\epsilon_\mu^*-\epsilon^* \cdot P(P+p)_\mu \right]  \nonumber\\
&&+T_3(q^2)\epsilon^* \cdot
P\left[q_\mu-\frac{q^2}{m^2_B-m_{K^*_2}^2}(P+p)_\mu \right] \, ,
\end{eqnarray}
where
\begin{eqnarray}
 A_3(q^2)&=&\frac{m_B+m_{K^*_2}}{2m_{K^*_2}}A_1(q^2)-\frac{m_B-m_{K^*_2}}{2m_{K^*_2}}A_2(q^2)\, ,
\end{eqnarray}
$A_0(0)=A_3(0)$,
$\epsilon^\alpha=\frac{\epsilon^{\alpha\beta}q_\beta}{m_B}$, and the
$\epsilon^{\alpha\beta}$ is the polarization vector of the tensor
meson $K_2^*(1430)$. In this article, we write down the calculations
for the $B\to K^*_2(1430)$ form-factors  explicitly, and obtain
others using  the flavor $SU(3)$ symmetry.

 We study the form-factors with the  two-point correlation functions $\Pi^k_{\alpha\beta\mu}(p,q)$,
\begin{eqnarray}
\Pi^k_{\alpha\beta\mu}(p,q)&=&i \int d^4x \, e^{i p \cdot x}
\langle 0 |T\left\{J_{\alpha\beta}(x) J^k_{\mu}(0)\right\}|B(P)\rangle \, ,\nonumber \\
J_{\alpha\beta}(x)&=&\frac{i}{2}\left\{ \bar{u}(x)\gamma_\alpha
\left[\overrightarrow{D}_\beta(x)-\overleftarrow{D}_\beta(x)\right]s(x)+\bar{u}(x)\gamma_\beta
\left[\overrightarrow{D}_\alpha(x)-\overleftarrow{D}_\alpha(x)\right]s(x)\right\}\, ,\\
J^1_\mu(x)&=&\bar{s}(x)\gamma_\mu b(x)\, ,\nonumber \\
J^2_\mu(x)&=&\bar{s}(x)\gamma_\mu\gamma_5 b(x)\, ,\nonumber \\
J^3_\mu(x)&=&\bar{s}(x)\sigma_{\mu\nu}q^\nu b(x)\, ,\nonumber \\
J^4_\mu(x)&=&\bar{s}(x)\sigma_{\mu\nu}\gamma_5q^\nu b(x)\, ,
\end{eqnarray}
 where $\overrightarrow{D}_\alpha(x)=\overrightarrow{\partial}_\alpha(x)-ig_sA_\alpha(x)$,
 $\overleftarrow{D}_\alpha(x)=\overleftarrow{\partial}_\alpha(x)+ig_sA_\alpha(x)$,
 $A_\alpha=\frac{\lambda^a}{2}A_\alpha^a$,  and $k=1,2,3,4$.

According to the quark-hadron duality \cite{SVZ79,Reinders85}, we
can insert  a complete set of intermediate states with the same
quantum numbers as the current operator  $J_{\alpha\beta}(x)$ into
the correlation functions $\Pi^k_{\alpha\beta\mu}(p,q) $ to obtain
the hadronic representation. After isolating the ground state
contributions from the pole term of the tensor meson $K_2^*(1430)$,
we obtain the results,
\begin{eqnarray}
\Pi^1_{\alpha\beta\mu}(p,q) &=&\frac{f_{K^*_2}m_{K^*_2}^2m_BV(q^2)}
  {\left(m_B+m_{K^*_2}\right)\left(m_{K^*_2}^2-p^2\right)}
  \left(-v_\alpha\varepsilon_{\beta\mu\lambda\tau}p^\lambda v^\tau
  -v_\beta\varepsilon_{\alpha\mu\lambda\tau}p^\lambda v^\tau\right)   + \cdots \, , \\
    Z^\alpha Z^\beta \Pi^2_{\alpha\beta\mu}(p,q) &=&\frac{if_{K^*_2}m_{K^*_2}^2}
  {m_{K^*_2}^2-p^2}
  \left[\left(m_B+m_{K^*_2}\right)A_1(q^2)Z_\mu Z\cdot v-\frac{m_B^2\widetilde{A}_2(q^2)}{m_B+m_{K^*_2}}(Z\cdot
  v)^2v_\mu\right.\nonumber\\
  &&\left.-\frac{m_B\widetilde{A}_3(q^2)}{m_B+m_{K^*_2}}(Z\cdot v)^2p_\mu \right]   + \cdots \, ,  \\
  \Pi^3_{\alpha\beta\mu}(p,q) &=&\frac{if_{K^*_2}m_{K^*_2}^2m_BT_1(q^2)}
  {m_{K^*_2}^2-p^2}  \left(-v_\alpha\varepsilon_{\beta\mu\lambda\tau}p^\lambda v^\tau
  -v_\beta\varepsilon_{\alpha\mu\lambda\tau}p^\lambda v^\tau\right)   + \cdots \, , \\
   Z^\alpha Z^\beta \Pi^4_{\alpha\beta\mu}(p,q) &=&\frac{f_{K^*_2}m_{K^*_2}^2}
  {m_{K^*_2}^2-p^2}
  \left[\left(m_B^2-m_{K^*_2}^2\right)T_2(q^2)Z_\mu Z\cdot v-m_B^2\widetilde{T}_3(q^2)(Z\cdot
  v)^2v_\mu \right]   + \cdots \, ,
 \end{eqnarray}
where
 \begin{eqnarray}
 \widetilde{A}_2(q^2)&=&A_2(q^2)+\frac{2m_{K^*_2}\left(m_B+m_{K^*_2}\right)}{q^2}\left[
 A_3(q^2)-A_0(q^2)\right] \, ,\nonumber\\
 \widetilde{A}_3(q^2)&=&A_2(q^2)-\frac{2m_{K^*_2}\left(m_B+m_{K^*_2}\right)}{q^2}\left[
 A_3(q^2)-A_0(q^2)\right]\, ,\nonumber\\
 \widetilde{T}_3(q^2)&=&T_2(q^2)-T_3(q^2)\left(1-
\frac{q^2}{m_B^2-m_{K^*_2}^2}\right)\, ,
 \end{eqnarray}
 the $Z_\mu$ is a four-vector with $Z^2=1$, $P_\alpha=p_\alpha+q_\alpha=m_Bv_\alpha$, and
   the  decay constant (or pole residue) $f_{K^*_2}$ is defined by
 \begin{eqnarray}
 \langle0|J_{\alpha\beta}(0)|K_2^*(p)\rangle &=& f_{K^*_2}m_{K^*_2}^2
 \epsilon_{\alpha\beta}\, , \nonumber\\
\sum_{\lambda}\epsilon^*_{\alpha\beta}(\lambda,p)\epsilon_{\mu\nu}(\lambda,p)
 &=&\frac{T_{\alpha\mu}T_{\beta\nu}+T_{\alpha\nu}T_{\beta\mu}}{2}-\frac{T_{\alpha\beta}T_{\mu\nu}}{3}\, , \nonumber \\
T_{\alpha\beta}&=&-g_{\alpha\beta}+\frac{p_\alpha p_\beta}{p^2} \, .
 \end{eqnarray}
 The tensor current $J_{\alpha\beta}(x)$ maybe also have non-vanishing
 coupling with the vector meson $K^*$,
 \begin{eqnarray}
\langle0|J_{\alpha\beta}(0)|K^*(p)\rangle &=& g_{K^*}m_{K^*}^2
 \left(\eta_{\alpha}p_\beta+\eta_{\beta}p_\alpha \right)\, ,
 \end{eqnarray}
 where the $\eta_\alpha$ is the polarization vector and the $g_{K^*}$ is the decay constant (or pole residue). In this
article, we derive the QCD sum rules with the tensor structures
$v_\alpha\varepsilon_{\beta\mu\lambda\tau}p^\lambda v^\tau
  +v_\beta\varepsilon_{\alpha\mu\lambda\tau}p^\lambda v^\tau$, $Z_\mu Z\cdot v$, $(Z\cdot v)^2v_\mu$ and
$(Z\cdot v)^2p_\mu$  to avoid possible
 contaminations from the $K^*$ meson.

 In the following, we briefly outline
the operator product expansion for the correlation functions
$\Pi^k_{\alpha\beta\mu}(p,q)$ in perturbative QCD. The calculations
are performed at the large space-like momentum region $p^2\ll 0$ and
$0 \leq q^2< m_b^2+\frac{m_b p^2}{\bar{\Lambda}}$,
$\bar{\Lambda}=m_B-m_b$. We write down the propagator of a massive
quark in the external gluon field in the Fock-Schwinger gauge and
the $B$-meson light-cone distribution amplitudes firstly
\cite{KhodjamirianB07,Grozin1997,Belyaev94},
\begin{eqnarray}
S_{ij}(x,y)&=&
 i \int\frac{d^4k}{(2\pi)^4}e^{-ik(x-y)}\left\{
\frac{\not\!k +m}{k^2-m^2} \delta_{ij} -\int\limits_0^1 du\, g_s \,
G^{\mu\nu}_{ij}(ux+(1-u)y)\right.\nonumber\\
 &&\left. \left[ \frac12 \frac {\not\!k
+m}{(k^2-m^2)^2}\sigma_{\mu\nu} - \frac1{k^2-m^2}u(x-y)_\mu
\gamma_\nu \right]\right\}\, ,
\end{eqnarray}

\begin{eqnarray}
 \langle 0|\bar{q}_{\alpha}(x)
h_{v\beta}(0) |B(v)\rangle &=& -\frac{if_B m_B}{4}\int\limits
_0^\infty d\omega e^{-i\omega v\cdot x}  \left \{(1 +\not\!v) \left
[ \phi_+(\omega) - \frac{\phi_+(\omega) -\phi_-(\omega)}{2 v\cdot
x}\not\! x \right ]\gamma_5\right\}_{\beta\alpha} \, , \nonumber
\end{eqnarray}

\begin{eqnarray}
\langle 0|\bar{q}_\alpha(x) G_{\lambda\rho}(ux)
h_{v\beta}(0)|B(v)\rangle &=& \frac{f_Bm_B}{4}\int\limits_0^\infty
d\omega \int\limits_0^\infty d\xi  e^{-i(\omega+u\xi) v\cdot
x}\left\{(1 + \not\!v) \Big[
(v_\lambda\gamma_\rho-v_\rho\gamma_\lambda)
 \right.
\nonumber \\
&&  \left(\Psi_A(\omega,\xi)-\Psi_V(\omega,\xi)\right)
-i\sigma_{\lambda\rho}\Psi_V(\omega,\xi)-\frac{x_\lambda
v_\rho-x_\rho v_\lambda}{v\cdot x}X_A(\omega,\xi)\nonumber\\
&&\left. +\frac{x_\lambda \gamma_\rho-x_\rho \gamma_\lambda}{v\cdot
x}Y_A(\omega,\xi)\Big]\gamma_5\right\}_{\beta\alpha}\,,
\end{eqnarray}
where
\begin{eqnarray}
\phi_+(\omega)&=&
\frac{\omega}{\omega_0^2}e^{-\frac{\omega}{\omega_0}} \, ,\,\,\,
\phi_-(\omega)= \frac{1}{\omega_0}e^{-\frac{\omega}{\omega_0}} \, ,\nonumber\\
 \Psi_A(\omega,\xi)& =& \Psi_V(\omega,\xi) = \frac{\lambda_E^2
}{6\omega_0^4}\xi^2 e^{-\frac{\omega+\xi}{\omega_0}} \, ,\nonumber\\
X_A(\omega,\xi)& = & \frac{\lambda_E^2 }{6\omega_0^4}\xi(2\omega-\xi)e^{-\frac{\omega+\xi}{\omega_0}}\,,\nonumber\\
Y_A(\omega,\xi)& =&  -\dfrac{\lambda_E^2 }{24\omega_0^4}
\xi(7\omega_0-13\omega+3\xi)e^{-\frac{\omega+\xi}{\omega_0}}\,,
\end{eqnarray}
the $\omega_0$ and $\lambda^2_E$ are  parameters of the $B$-meson
light-cone distribution amplitudes. In this article, we take the
two-particle and three-particle $B$-meson light-cone distribution
amplitudes suggested in Ref.\cite{Grozin1997} and
Ref.\cite{KhodjamirianB07}, respectively. They obey the powerful
constraints derived in Ref.\cite{Qiao2001} and the relations between
the matrix elements of the local operators and the moments of the
light-cone distribution amplitudes,   if the conditions $
\omega_0=\frac{2}{3} \bar{\Lambda}$ and
$\lambda_E^2=\lambda_H^2=\frac{3}{2}\omega_0^2= \frac{2}{3}
\bar{\Lambda}^2$
 are satisfied \cite{Grozin1997}.

 \begin{figure}
 \centering
 \includegraphics[totalheight=4cm,width=14cm]{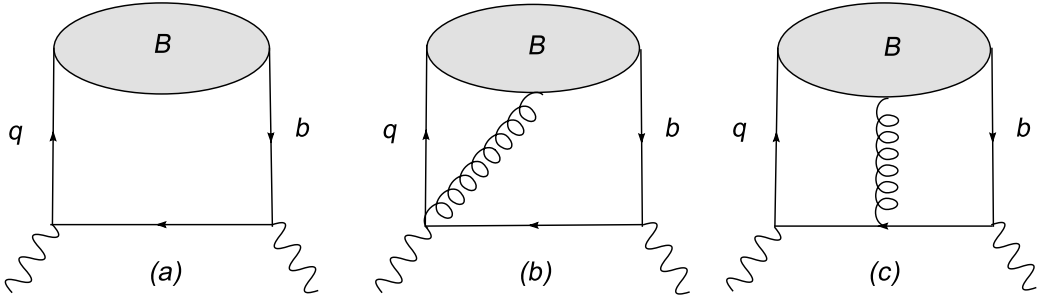}
    \caption{ The diagrams contribute to the form-factors in the operator product expansion.  }
\end{figure}

We contract the quark fields with Wick theorem, substitute the $s$
quark propagator and the  $B$-meson light-cone distribution
amplitudes into the correlation functions
$\Pi^k_{\alpha\beta\mu}(p,q)$, then complete the integrals over the
variables $x$ and $k$, finally obtain the representation at the
level of quark-gluon degrees of freedom. In calculations, we take
into account the contributions from the two-particle and
three-particle $B$-meson light-cone distribution amplitudes, the
gluons presented in the covariant derivatives
$\overrightarrow{D}_\alpha$, $\overleftarrow{D}_\alpha$ (or emitted
from the vertex) and emitted from the intermediate quark lines,
which correspond to the diagrams (b) and (c) respectively in Fig.1,
both contribute to the three-particle $B$-meson light-cone
distribution amplitudes. In calculating the diagrams (b-c) in Fig.1,
we use the Fock-Schwinger gauge $x^\mu A_\mu^a(x)=0$ to express  the
gluon field $A_\mu $ in terms of the gluon field strength tensor
$G_{\mu\nu}$, $A_\mu(x)=\int_0^1 d\tau \tau x_\nu G_{\nu\mu}(\tau
x)$,  and then extract the three-particle $B$-meson light-cone
distribution amplitudes.

In the region of small $\omega$, the exponential form of (or the
Gaussian-like)  distribution amplitude $\phi_+(\omega)$ is
numerically close to the more elaborated model (the
Braun-Ivanov-Korchemsky model, or the BIK model) suggested in
Ref.\cite{Braun2004},
\begin{eqnarray}
\phi_+(\omega, \mu=1 \mbox{GeV}) =\frac{4\omega}{\pi
\lambda_B(1+\omega^2)} \left[\frac{1}{1+\omega^2}-2
\frac{\sigma_B-1}{\pi^2} \ln\omega\right]\, ,
 \end{eqnarray}
 where $\omega_0=\lambda_B$, and $\omega$ is in unit of GeV. The parameters $\lambda_B$ and
$\sigma_B$ are determined by the QCD sum rules  including the
radiative and nonperturbative corrections in the heavy quark
effective theory. There are other phenomenological models for the
two-particle $B$-meson light-cone distribution amplitudes, for
example, the $k_T$ factorization formalism \cite{Bdis1,Bdis2}. In
this article, we use the QCD sum rules motivated models.

 After matching  with the hadronic representation below the continuum thresholds $s_0$, we
obtain the sum rules for the form-factors $V(q^2)$, $A_1(q^2)$,
$\widetilde{A}_2(q^2)$, $\widetilde{A}_3(q^2)$, $T_1(q^2)$,
$T_2(q^2)$ and $\widetilde{T}_3(q^2)$, respectively,

\begin{eqnarray}
V(q^2) &=& \frac{f_Bm_B(m_B+m_{K^*_2})}{f_{K^*_2}
m_{K^*_2}^2}e^{\frac{m_{K^*_2}^2}{M^2}}\left\{
\int_{0}^{\sigma_0}d\sigma\omega'
\left\{\frac{\phi_+(\omega')}{\bar{\sigma}}-\frac{m_s}{\bar{\sigma}^2M^2}
\left[\widetilde{\phi}_+(\omega')-\widetilde{\phi}_-(\omega')\right] \right\}e^{-\frac{s}{M^2}}  \right.\nonumber\\
&&+\int_{0}^{\sigma_0}\widetilde{d\sigma}\left\{
\frac{\left[2um_s+(2u-1)\omega\right]\left[\Psi_A(\omega,\xi)-\Psi_V(\omega,\xi)\right]}{\bar{\sigma}^2M^2}
-\frac{2\left[u\widetilde{\omega}+(1-u)\omega\right]\Psi_V(\omega,\xi)}{\bar{\sigma}^2M^2}\right.  \nonumber \\
&&+
\frac{2u(\widetilde{\omega}+\omega)X_A(\omega,\xi)}{\bar{\sigma}^2M^2}
+\frac{2m_s(\omega-2u\widetilde{\omega})\widetilde{Y}(\omega,\xi)}{\bar{\sigma}^3M^4}
+\frac{\widetilde{X}_A(\omega,\xi)}{\bar{\sigma}^2M^2} \left[
\frac{1+8u}{2} \right.\nonumber \\
&&\left.\left.
\left.+\frac{2(2um_s^2-m_s\omega+\omega\widetilde{\omega})}{\bar{\sigma}M^2}
+\frac{2\omega}{m_B\bar{\sigma}}\left(
1-\frac{\widetilde{m}_B^2}{2M^2}\right)\right] \right\}
e^{-\frac{s}{M^2}}\right\} \, ,
\end{eqnarray}

\begin{eqnarray}
A_1(q^2) &=& \frac{2f_Bm_B^2}{f_{K^*_2}
m_{K^*_2}^2(m_B+m_{K^*_2})}e^{\frac{m_{K^*_2}^2}{M^2}}\left\{
\int_{0}^{\sigma_0}d\sigma\omega'
\left\{\frac{\phi_+(\omega')}{\bar{\sigma}}\left(m_s-\omega'+\frac{\widetilde{m}_B^2}{2m_B}\right)
\right.\right.\nonumber \\
&&\left.-\left[\widetilde{\phi}_+(\omega')-\widetilde{\phi}_-(\omega')\right]\frac{m_s}{\bar{\sigma}^2}
\left[ \frac{m_s-\omega'}{M^2}-\frac{1}{2m_B}\left(1-\frac{\widetilde{m}_B^2}{M^2}\right)\right]\right\}e^{-\frac{s}{M^2}}  \nonumber\\
&&+\int_{0}^{\sigma_0}\widetilde{d\sigma}\left\{
\left[\Psi_A(\omega,\xi)-\Psi_V(\omega,\xi)\right]\left[-\frac{u}{\bar{\sigma}}
+\frac{2um_s(m_s-\widetilde{\omega})+\omega(m_s+\widetilde{\omega})-2u\omega\widetilde{\omega}}{\bar{\sigma}^2M^2}\right.
\right.  \nonumber \\
&&\left.-\frac{2u(m_s+\omega)-\omega}{2m_B\bar{\sigma}^2}\left(1-\frac{\widetilde{m}_B^2}{M^2}
\right) \right] +\frac{\Psi_V(\omega,\xi)}{\bar{\sigma}^2}\left[
\frac{2\left[u\widetilde{\omega}(\widetilde{\omega}-m_s)+(1-u)\omega\widetilde{\omega}\right]}{M^2}\right.
\nonumber \\
&&\left.+\frac{u(\widetilde{\omega}-\omega)+\omega}{m_B}\left(1-\frac{\widetilde{m}_B^2}{M^2}
\right) \right]+\frac{2uX_A(\omega,\xi)}{\bar{\sigma}^2} \left[
\frac{\widetilde{\omega}(m_s-\widetilde{\omega}-\omega)}{M^2}-\frac{\omega+\widetilde{\omega}}{2m_B}\left(1-\frac{\widetilde{m}_B^2}{M^2}
\right) \right]
\nonumber \\
&&+\frac{\widetilde{X}_A(\omega,\xi)}{\bar{\sigma}^2} \left[
\frac{m_s+6\omega+(1-12u)\widetilde{\omega}}{2M^2}
-\frac{1+8u}{4m_B}\left(1-\frac{\widetilde{m}_B^2}{M^2}\right)
+\frac{4um_s^2(m_s-\widetilde{\omega}-\omega)}{\bar{\sigma}M^4} \right.\nonumber \\
&&\left.-\frac{4um_s^2}{\bar{\sigma}m_BM^2}\left(1-\frac{\widetilde{m}_B^2}{2M^2}\right)
-\frac{4\omega}{\bar{\sigma}M^2}\left(1-\frac{s}{2M^2}
\right)-\frac{2\omega}{\bar{\sigma}m_B^2} \left(
\frac{1}{2}-\frac{\widetilde{m}_B^2}{M^2}+\frac{\widetilde{m}_B^4}{4M^4}\right)\right]\nonumber \\
&&+\frac{2\widetilde{Y}_A(\omega,\xi)}{\bar{\sigma}^2M^2}
\left[(2u-1)(\widetilde{\omega}+\omega)+2u\widetilde{\omega}
+\frac{m_s\left[2u\widetilde{\omega}(\widetilde{\omega}-m_s)-\omega(\widetilde{\omega}+m_s)+2um_s\omega\right]}{\bar{\sigma}M^2}
\right.\nonumber \\
&&\left.\left.\left.
+\frac{m_s(2u\widetilde{\omega}-\omega)}{\bar{\sigma}m_B}\left(1-\frac{\widetilde{m}_B^2}{2M^2}
\right)\right]\right\}e^{-\frac{s}{M^2}} \right\}\, ,
\end{eqnarray}

\begin{eqnarray}
\widetilde{A}_2(q^2) &=& -\frac{2f_B(m_B+m_{K^*_2})}{f_{K^*_2}
m_{K^*_2}^2}e^{\frac{m_{K^*_2}^2}{M^2}}\left\{
\int_{0}^{\sigma_0}d\sigma{\omega'}^2
\left\{\frac{2\phi_+(\omega')}{\bar{\sigma}}
 +\frac{m_s}{\bar{\sigma}^2M^2}\left[\widetilde{\phi}_+(\omega')-\widetilde{\phi}_-(\omega')\right]
\right\}e^{-\frac{s}{M^2}} \right. \nonumber\\
&&+\int_{0}^{\sigma_0}\widetilde{d\sigma}\left\{
\frac{2\left[(1-2u)\omega\widetilde{\omega} -2m_s\omega-2u\widetilde{\omega}^2  \right]\left[\Psi_A(\omega,\xi)-\Psi_V(\omega,\xi)\right]}{\bar{\sigma}^2M^2}   \right.\nonumber \\
&&+\frac{4\widetilde{\omega}\left[(1-u)\omega-u\widetilde{\omega}\right]\Psi_V(\omega,\xi)}{\bar{\sigma}^2M^2}
+\frac{4u\widetilde{\omega}(\omega+\widetilde{\omega})X_A(\omega,\xi)}{\bar{\sigma}^2M^2}
\nonumber \\
&&+\frac{\widetilde{X}_A(\omega,\xi)}{\bar{\sigma}^2M^2} \left[
(12u-1)\widetilde{\omega}-6\omega+\frac{8um_s^2(\omega+\widetilde{\omega})}{\bar{\sigma}M^2}
-\frac{4\omega\widetilde{\omega}}{\bar{\sigma}m_B}\left(1-\frac{\widetilde{m}_B^2}{2M^2}\right)
 \right.\nonumber \\
&&\left.\left.\left.
+\frac{8\omega}{\bar{\sigma}}\left(1-\frac{s}{2M^2} \right)\right]
+\frac{4\widetilde{Y}_A(\omega,\xi)}{\bar{\sigma}^3M^4}
\left[\omega\widetilde{\omega}(2m_s-\widetilde{\omega})+2u(\omega+\widetilde{\omega})\widetilde{\omega}^2
\right] \right\}e^{-\frac{s}{M^2}} \right\}\, ,
\end{eqnarray}

\begin{eqnarray}
\widetilde{A}_3(q^2) &=& \frac{2f_Bm_B(m_B+m_{K^*_2})}{f_{K^*_2}
m_{K^*_2}^2}e^{\frac{m_{K^*_2}^2}{M^2}}\left\{
\int_{0}^{\sigma_0}d\sigma\omega'
\left\{\frac{\phi_+(\omega')}{\bar{\sigma}}
 +\frac{2\omega'-m_s}{\bar{\sigma}^2M^2}\left[\widetilde{\phi}_+(\omega')-\widetilde{\phi}_-(\omega')\right]
\right\}e^{-\frac{s}{M^2}} \right. \nonumber\\
&&+\int_{0}^{\sigma_0}\widetilde{d\sigma}\left\{
\frac{\left[\Psi_A(\omega,\xi)-\Psi_V(\omega,\xi)\right]\left[2u(m_s-2\widetilde{\omega}-\omega)-\omega \right]}{\bar{\sigma}^2M^2}
-\frac{2\Psi_V(\omega,\xi)\left[(1-u)\omega+u\widetilde{\omega}\right]}{\bar{\sigma}^2M^2} \right.\nonumber \\
&&+\frac{2uX_A(\omega,\xi)(\omega+\widetilde{\omega})}{\bar{\sigma}^2M^2}
+\frac{\widetilde{X}_A(\omega,\xi)}{\bar{\sigma}^2M^2}
\left[\frac{1}{2}+4u+
\frac{2\omega(m_s+\widetilde{\omega})+4um_s^2}{\bar{\sigma}M^2}
+\frac{2\omega}{\bar{\sigma}m_B}\left(1-\frac{\widetilde{m}_B^2}{2M^2}\right)
 \right]\nonumber \\
&&\left.\left.
+\frac{2\widetilde{Y}_A(\omega,\xi)}{\bar{\sigma}^3M^4}
\left[\omega(m_s-2\widetilde{\omega})+2u\widetilde{\omega}(2\widetilde{\omega}-m_s+2\omega)
\right] \right\}e^{-\frac{s}{M^2}} \right\}\, ,
\end{eqnarray}

\begin{eqnarray}
T_1(q^2) &=& -\frac{f_Bm_B}{f_{K^*_2}
m_{K^*_2}^2}e^{\frac{m_{K^*_2}^2}{M^2}}\left\{
\int_{0}^{\sigma_0}d\sigma\omega'(\omega'-m_B-m_s)
\left\{\frac{\phi_+(\omega')}{\bar{\sigma}}-\frac{m_s}{\bar{\sigma}^2M^2} \left[\widetilde{\phi}_+(\omega')-\widetilde{\phi}_-(\omega')\right]\right\}  \right.\nonumber\\
&&e^{-\frac{s}{M^2}}+\int_{0}^{\sigma_0}\widetilde{d\sigma}\left\{\left[\frac{u}{\bar{\sigma}}+
\frac{\left[\omega(m_B-\widetilde{\omega}-m_s)-2um_s(m_s+m_B-\widetilde{\omega})+2u\omega(\widetilde{\omega}-m_B)\right]
}{\bar{\sigma}^2M^2}\right] \right. \nonumber \\
&&\left[\Psi_A(\omega,\xi)-\Psi_V(\omega,\xi)\right]
+\frac{2\left[(1-u)\omega(m_B-\widetilde{\omega})+u\widetilde{\omega}(m_B+m_s-\widetilde{\omega})
\right]\Psi_V(\omega,\xi)}{\bar{\sigma}^2M^2}\nonumber\\
&& +
\frac{2u\left[\widetilde{\omega}(\widetilde{\omega}-m_B-m_s)+\omega(\widetilde{\omega}-m_B)\right]X_A(\omega,\xi)}{\bar{\sigma}^2M^2}
+\frac{\widetilde{X}_A(\omega,\xi)}{\bar{\sigma}^2M^2} \left[- \frac{6\omega+\widetilde{\omega}+m_B+m_s}{2}\right. \nonumber \\
&&
 +4u(\widetilde{\omega}-m_B)+2u\widetilde{\omega}+\frac{2\left[m_B\omega(m_s-\widetilde{\omega})+2um_s^2(\widetilde{\omega}-m_B-m_s+\omega)\right]}{\bar{\sigma}M^2}
\nonumber\\
&&\left.+\frac{4\omega}{\bar{\sigma}}\left(1-\frac{s}{2M^2}\right)-\frac{2\omega(\widetilde{\omega}+m_B-m_s)}{m_B\bar{\sigma}}\left(
1-\frac{\widetilde{m}_B^2}{2M^2}\right)\right]+\frac{2\widetilde{Y}_A(\omega,\xi)}{\bar{\sigma}^2M^2}\left[
(1-2u)\omega
\right.\nonumber\\
&&\left.\left.\left.+(1-4u)\widetilde{\omega}
+\frac{m_s\left[\omega(m_s+\widetilde{\omega}-m_B)+2u\widetilde{\omega}(m_s-\widetilde{\omega}+m_B)-2um_s\omega\right]}{\bar{\sigma}M^2}\right]\right\}e^{-\frac{s}{m^2}}\right\}\,
,
\end{eqnarray}

\begin{eqnarray}
T_2(q^2) &=& \frac{2f_Bm_B^2}{f_{K^*_2}
m_{K^*_2}^2(m_B^2-m_{K^*_2}^2)}e^{\frac{m_{K^*_2}^2}{M^2}}\left\{
\int_{0}^{\sigma_0}d\sigma\omega'
\left\{\frac{\phi_+(\omega')}{\bar{\sigma}}\left[m_B(m_s-\omega') -s
+\frac{\widetilde{m}_B^2(\omega'+m_B-m_s)}{2m_B}\right]
\right.\right.
\nonumber\\
&&\left.+\frac{m_s}{\bar{\sigma}^2}
\left[\frac{m_B(\omega'-m_s)}{M^2}+\frac{m_B+\omega'-m_s}{2m_B}\left(1-\frac{\widetilde{m}_B^2}{M^2}\right) -\left(1-\frac{s}{M^2}\right)\right]
\left[\widetilde{\phi}_+(\omega')-\widetilde{\phi}_-(\omega')\right] \right\}  \nonumber\\
&&e^{-\frac{s}{M^2}}+\int_{0}^{\sigma_0}\widetilde{d\sigma}\left\{\left[\Psi_A(\omega,\xi)-\Psi_V(\omega,\xi)\right]
\left[-\frac{u}{\bar{\sigma}}\left(m_B-\frac{\widetilde{m}_B^2}{2m_B}\right) \right.\right. \nonumber \\
&&
+\frac{m_B\left[2um_s(m_s-\widetilde{\omega})+\omega(m_s+\widetilde{\omega})-2u\omega\widetilde{\omega}\right]
}{\bar{\sigma}^2M^2}+\frac{2u(m_s+\omega)-\omega}{\bar{\sigma}^2}\left(1-\frac{s}{M^2}\right) \nonumber\\
&&\left. +\frac{\omega(\widetilde{\omega}+m_B+m_s)-2um_s(\widetilde{\omega}+m_B-m_s)-2u\omega(m_B+\widetilde{\omega})}{2m_B\bar{\sigma}^2}\left(1-\frac{\widetilde{m}_B^2}{M^2}\right)\right] \nonumber\\
&&+\frac{2\Psi_V(\omega,\xi)}{\bar{\sigma}^2} \left[
\frac{m_B\left[(1-u)\omega\widetilde{\omega}+u\widetilde{\omega}(\widetilde{\omega}-m_s)\right]}{M^2}
-\left[(1-u)\omega+u\widetilde{\omega}\right]\left(1-\frac{s}{M^2}\right)\right.\nonumber\\
&&\left.+\frac{(1-u)\omega(m_B+\widetilde{\omega})+u\widetilde{\omega}(\widetilde{\omega}+m_B-m_s)}{2m_B}
\left(1-\frac{\widetilde{m}_B^2}{M^2}\right)\right]+\frac{2uX_A(\omega,\xi)}{\bar{\sigma}^2}\nonumber\\
 && \left[ \frac{m_B\widetilde{\omega}(m_s-\widetilde{\omega}-\omega)}{M^2}
 -\frac{\widetilde{\omega}(\widetilde{\omega}+m_B-m_s)+\omega(\widetilde{\omega}+m_B)}{2m_B}\left(1-\frac{\widetilde{m}_B^2}{M^2}\right)
 +(\widetilde{\omega}+\omega)\right.\nonumber\\
&&\left.\left(1-\frac{s}{M^2}\right)\right]
+\frac{\widetilde{X}_A(\omega,\xi)}{\bar{\sigma}^2} \left[
\frac{m_B(m_s+6\omega+\widetilde{\omega})-12um_B\widetilde{\omega}}{2M^2}+
\frac{1+8u}{2}\left( 1-\frac{s}{M^2}\right)\right. \nonumber \\
&&
+\frac{4um_s^2m_B(m_s-\widetilde{\omega}-\omega)}{\bar{\sigma}M^4}
+\frac{6\omega+\widetilde{\omega}+m_s-m_B-8u(\widetilde{\omega}+m_B)-4u\widetilde{\omega}}{4m_B}\left(1-\frac{\widetilde{m}_B^2}{M^2}\right)
\nonumber\\
&&\left.+\frac{8um_s^2-4\omega(m_B+m_s-\widetilde{\omega})}{\bar{\sigma}M^2}\left(1-\frac{s}{2M^2}\right)
-\frac{4um_s^2(\widetilde{\omega}+m_B-m_s+\omega)}{m_B\bar{\sigma}M^2}\left(
1-\frac{\widetilde{m}_B^2}{2M^2}\right)\right.\nonumber\\
&&\left.-\frac{2\omega(m_B+m_s-\widetilde{\omega})}{m_B^2\bar{\sigma}}\left(\frac{1}{2}-\frac{\widetilde{m}_B^2}{M^2}+
\frac{\widetilde{m}_B^4}{4M^4}\right)\right]+\frac{2\widetilde{Y}_A(\omega,\xi)}{\bar{\sigma}^2}\left[
\frac{m_B\left[2u\widetilde{\omega}+(2u-1)(\omega+\widetilde{\omega})\right]}{M^2}
\right.\nonumber\\
&&+\frac{(2u-1)(\omega+\widetilde{\omega})+2u\widetilde{\omega}}{2m_B}\left(1-\frac{\widetilde{m}_B^2}{M^2}\right)
+\frac{m_sm_B\left[2u\widetilde{\omega}(\widetilde{\omega}-m_s)-\omega(m_s+\widetilde{\omega})+2um_s\omega\right]}{\bar{\sigma}M^4}
\nonumber\\
&&+\frac{m_s\left[2u\widetilde{\omega}(m_B+\widetilde{\omega}-m_s)-\omega(\widetilde{\omega}+m_B+m_s)+2um_s\omega\right]}{m_B\bar{\sigma}M^2}\left(1-\frac{\widetilde{m}_B^2}{2M^2}\right)
\nonumber\\
&&\left.\left.\left.+\frac{2m_s(\omega-2u\widetilde{\omega})}{\bar{\sigma}M^2}\left(1-\frac{s}{2M^2}\right)\right]\right\}e^{-\frac{s}{M^2}}
\right\} \, ,
\end{eqnarray}

\begin{eqnarray}
\widetilde{T}_3(q^2) &=& -\frac{2f_B}{f_{K^*_2}
m_{K^*_2}^2}e^{\frac{m_{K^*_2}^2}{M^2}}\left\{
\int_{0}^{\sigma_0}d\sigma\omega'
\left\{\frac{\phi_+(\omega')m_B(\omega'-m_s)}{\bar{\sigma}}+\frac{1}{\bar{\sigma}}\left[\widetilde{\phi}_+(\omega')-\widetilde{\phi}_-(\omega')\right]
\left[m_B  \right.\right.\right.\nonumber\\
&&\left.\left.+\frac{m_B\omega'(\omega'-m_s)}{\bar{\sigma}M^2}
+\frac{2\omega'-m_B}{\bar{\sigma}}\left(1-\frac{s}{M^2}\right)\right]
\right\} e^{-\frac{s}{M^2}}+\int_{0}^{\sigma_0}\widetilde{d\sigma}
\left\{\left[\Psi_A(\omega,\xi)-\Psi_V(\omega,\xi)\right] \right. \nonumber\\
&&\left[-\frac{um_B}{\bar{\sigma}}+
\frac{m_B\left[2u\widetilde{\omega}(m_s-\widetilde{\omega})-\omega(m_s+\widetilde{\omega})+2u\omega\widetilde{\omega}\right]
}{\bar{\sigma}^2M^2}-\frac{2u\left[(2\widetilde{\omega}-m_B)+2\omega\right]}{\bar{\sigma}^2}\right.\nonumber \\
&&\left.\left(1-\frac{s}{M^2}\right)
+\frac{2u\omega}{\bar{\sigma}^2}\left(1-\frac{\widetilde{m}_B^2}{M^2}\right)\right]
+\frac{2m_B\widetilde{\omega}\left[u(m_s-\widetilde{\omega})+(u-1)\omega\right]\Psi_V(\omega,\xi)}{\bar{\sigma}^2M^2}
\nonumber\\
&& +\frac{2u\left[
m_B\widetilde{\omega}(\widetilde{\omega}-m_s+\omega)\right]X_A(\omega,\xi)}{\bar{\sigma}^2M^2}
+\frac{\widetilde{X}_A(\omega,\xi)}{\bar{\sigma}^2M^2}\left[
-\frac{m_B(6\omega+\widetilde{\omega}+m_s-12u\widetilde{\omega})}{2}\right.
\nonumber\\
&&\left.+\frac{4um_s^2m_B(\widetilde{\omega}-m_s+\omega)}{\bar{\sigma}M^2}+
\frac{4\omega(m_B+2m_s)}{\bar{\sigma}}\left(
 1-\frac{s}{2M^2}\right)-\frac{2m_B\omega(m_s+\widetilde{\omega})}{\bar{\sigma}m_B}\left(1-\frac{\widetilde{m}_B^2}{2M^2}\right)\right] \nonumber \\
&& +\frac{2\widetilde{Y}_A(\omega,\xi)}{\bar{\sigma}^2M^2}\left[
m_B\left[(1-2u)(\omega+\widetilde{\omega})-2u\widetilde{\omega}\right]+\frac{m_Bm_s\left[\omega(m_s+\widetilde{\omega})
+2u\widetilde{\omega}(m_s-\widetilde{\omega})-2um_s\omega
\right]}{\bar{\sigma}M^2}
\right.\nonumber\\
&&\left.\left.\left.-\frac{2\widetilde{\omega}\left[2u\widetilde{\omega}+(2u-1)\omega\right]}{\bar{\sigma}}\left(1-\frac{\widetilde{m}_B^2}{2M^2}\right)
+\frac{4\widetilde{\omega}\left[2u\widetilde{\omega}+(2u-1)\omega\right]}{\bar{\sigma}}\left(1-\frac{s}{2M^2}\right)\right]\right\}e^{-\frac{s}{M^2}}
\right\} \, ,\nonumber\\
\end{eqnarray}
where
\begin{eqnarray}
\int_{0}^{\sigma_0}\widetilde{d\sigma}&=&\int_{0}^{\sigma_0}d\sigma
\int_0^{\sigma m_B}d\omega \int_{\sigma m_B-\omega}^\infty
\frac{d\xi}{\xi} \, ,\nonumber\\
s &=& m_B^2 \sigma-\frac{\sigma q^2-m_s^2}{\bar{\sigma}}\, , \,\,\,\omega'=\sigma m_B \, , \, \, \,\bar{\sigma}=1-\sigma \, , \, \, \, \widetilde{\omega} =\omega+u\xi\, ,\nonumber\\
\sigma_0&=&\frac{s^{K^*_2}_0+m_B^2-q^2-\sqrt{(s^{K^*_2}_0+m_B^2-q^2)^2-4(s^{K^*_2}_0-m_s^2)m_B^2}}{2m_B^2} \, ,\nonumber\\
u&=&\frac{\sigma m_B-\omega}{\xi} \, ,\,\,\,\widetilde{m}_B^2=m_B^2(1+\sigma)-\frac{q^2-m_s^2}{\bar{\sigma}}\, ,\nonumber\\
\widetilde{X}_A(\omega,\xi)&=&\int_0^\omega d\lambda
X_A(\lambda,\xi) \, ,\,\,\,\widetilde{Y}_A(\omega,\xi)=\int_0^\omega
d\lambda
Y_A(\lambda,\xi) \, ,\nonumber\\
\widetilde{\phi}_\pm(\omega)&=&\int_0^\omega d\lambda
\phi_\pm(\lambda) \, .
\end{eqnarray}
With a simple replacement,
\begin{eqnarray}
 m_s&\to& 0\, , \, \, m_{K^*_2}\to m_{a_2}\, , \,\, f_{K^*_2}\to
 f_{a_2}\, , \,\, s_0^{K^*_2} \to s_0^{a_2} \, , \nonumber \\
 m_s&\to& 0\, , \, \, m_{K^*_2}\to m_{f_2}\, , \,\, f_{K^*_2}\to
 f_{f_2}\, , \,\, s_0^{K^*_2} \to s_0^{f_2} \, ,
\end{eqnarray}
we can obtain the corresponding sum rules for the $B\to a_2(1320)$
and $B\to f_2(1270)$ form-factors,  respectively.

In Ref.\cite{Lange2003}, Lange and  Neubert observe that the
evolution effects drive the light-cone distribution amplitude
$\phi_+(\omega)$ toward a linear growth at the origin and generate a
radiative tail that  falls off slower than $\frac{1}{\omega}$, even
if the initial function has an arbitrarily rapid falloff. The
normalization integral of the $\phi_{+}(\omega)$ is ultraviolet
divergent. In this article, we derive the sum rules without the
radiative $\mathcal {O}(\alpha_s)$ corrections, the ultraviolet
behavior of the $\phi_+(\omega)$ plays no role at the leading order
$\mathcal {O}(1)$. Furthermore, the duality thresholds in the sum
rules  are well below the region where the effect of the tail
becomes noticeable. The nontrivial renormalization of the $B$-meson
light-cone distribution amplitudes  is so far known only for the
$\phi_{+}(\omega)$,  we use the light-cone distribution amplitudes
of order $\mathcal {O}(1)$, which satisfy
 all QCD constraints.

\section{Numerical result and discussion}
The input parameters for the $B$-meson light-cone distribution
amplitudes are taken as
$\omega_0=\lambda_B(\mu)=(0.46\pm0.11)\,\rm{GeV}$, $\mu=1\,\rm{GeV}$
\cite{Braun2004}, $\lambda_E^2=(0.11\pm0.06)\,\rm{GeV}^2$
\cite{Grozin1997}, $m_B=5.279 \,\rm{GeV}$, and  $f_B=(0.18 \pm
0.02)\,\rm{GeV}$ \cite{LCSRreview}.

The masses, decay constants (or pole residues), threshold parameters
and Borel parameters of the tensor mesons $f_2(1270)$, $a_2(1320)$
and $K^*_2(1430)$ are determined by the conventional two-point QCD
sum rules. The values are $m_{K^*_2}=(1.43\pm0.01)\,\rm{GeV}$,
$m_{a_2}=(1.31\pm0.01)\,\rm{GeV}$,
$m_{f_2}=(1.27\pm0.01)\,\rm{GeV}$,
$f_{K^*_2}=(0.118\pm0.005)\,\rm{GeV}$,
$f_{a_2}=(0.107\pm0.006)\,\rm{GeV}$,
$f_{f_2}=(0.102\pm0.006)\,\rm{GeV}$, $s_0^{K^*_2}=3.13\,\rm{GeV}^2$,
$s_0^{a_2}=2.70\,\rm{GeV}^2$, $s_0^{f_2}=2.53\,\rm{GeV}^2$,
$M^2_{K_2^*}=(1.2-1.6)\,\rm{GeV}^2$,
$M^2_{a_2}=(1.0-1.4)\,\rm{GeV}^2$ and
 $M_{f_2}^2=(1.0-1.4)\,\rm{GeV}^2$ \cite{YangTLC}. The tensor meson
$K^*_2(1430)$ was originally studied with the QCD sum rules by T. M.
Aliev et al \cite{Aliev1430}. In the conventional two-point QCD sum
rules for the tensor mesons, we determine the Borel windows by
imposing the two criteria (pole dominance and convergence of the
operator product expansion) to reproduce the experimental values of
the masses \cite{YangTLC}. In this article, we take the tensor
currents $J_{\mu\nu}(x)$ to interpolate the tensor mesons
$K_2^*(1430)$, $a_2(1320)$ and $f_2(1270)$ as in the conventional
two-point QCD sum rules,  and expect that the parameters  survive.

\begin{figure}
\centering
\includegraphics[totalheight=7cm,width=10cm]{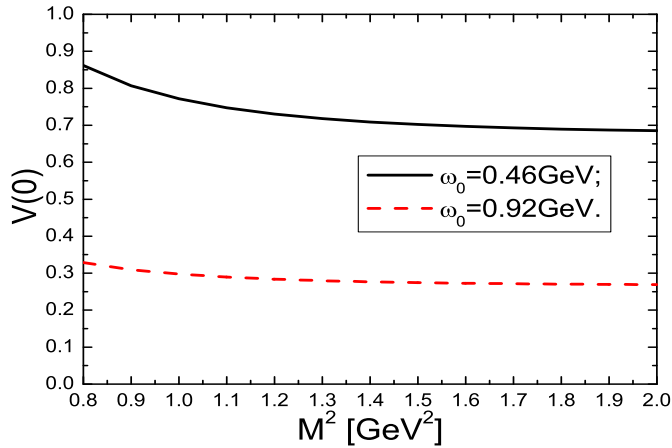}
  \caption{ The form-factor $V_{BK_2^*}(0)$ with variation of the  Borel parameter $M^2$,
  other parameters are taken to be the central values.  }
\end{figure}

Taking into account all uncertainties of the relevant parameters, we
obtain the numerical values of the  form-factors $V(q^2)$,
$A_1(q^2)$, $\widetilde{A}_2(q^2)$, $\widetilde{A}_3(q^2)$,
$T_1(q^2)$, $T_2(q^2)$ and $\widetilde{T}_3(q^2)$. The values of the
form-factors $V(0)$, $A_1(0)$, $\widetilde{A}_2(0)$,
$\widetilde{A}_3(0)$, $T_1(0)$, $T_2(0)$ are very stable with
variations of the Borel parameters in  a large range,   the
uncertainties originate from the Borel parameters are small (for
example, see Fig.2); while the values of the form-factors
$\widetilde{T}_3(0)$ are not stable enough with the variation of the
Borel parameter even in the Borel window.  The form-factors can be
parameterized  in the double-pole form,
 \begin{eqnarray}
 F_i(q^2)&=&\frac{F_i(0)}{1+a_Fq^2/m_B^2+b_Fq^4/m_B^4}  \, ,
 \end{eqnarray}
where the $F_i(q^2)$ denote the $V(q^2)$, $A_1(q^2)$,
$\widetilde{A}_2(q^2)$, $\widetilde{A}_3(q^2)$, $T_1(q^2)$,
$T_2(q^2)$ and $\widetilde{T}_3(q^2)$,  the $a_F$ and $b_F$ are the
corresponding coefficients. The values of the form-factors at zero
momentum transfer and the $a_F$, $b_F$ are presented in Table 1. In
Table 2, we present the central values of the ratios among the
form-factors at zero momentum transfer.

In this article,  we calculate the uncertainties $\delta$ with the
formula
\begin{eqnarray}
\delta=\sqrt{\sum_i\left(\frac{\partial f}{\partial
x_i}\right)^2\mid_{x_i=\bar{x}_i} (x_i-\bar{x}_i)^2}\,  ,
\end{eqnarray}
 where the $f$ denotes  the
$B\to T$ form-factors,  the $x_i$ denotes the input  parameters
$\omega_0$, $\lambda_E^2$, $f_{K^*_2}$, $\cdots$. As the partial
 derivatives   $\frac{\partial f}{\partial x_i}$ are difficult to carry
out analytically, we take the  approximation $\left(\frac{\partial
f}{\partial x_i}\right)^2 (x_i-\bar{x}_i)^2\approx
\left[f(\bar{x}_i\pm \Delta x_i)-f(\bar{x}_i)\right]^2$ in the
numerical calculations, and take into account the values
$\left[f(\bar{x}_i+ \Delta x_i)-f(\bar{x}_i)\right]^2
\neq\left[f(\bar{x}_i- \Delta x_i)-f(\bar{x}_i)\right]^2$.

\begin{table}
\begin{center}
\begin{tabular}{|cc|cc|cc|cc|}\hline\hline
                        &$\omega_0$               & $F_{BK^*_2}(0)$        &$[a_F,b_F]$        & $F_{Ba_2}(0)$          &$[a_F,b_F]$      & $F_{Bf_2}(0)$          &$[a_F,b_F]$ \\ \hline
    $V$                 &$\widetilde{\lambda}_B$  & $0.71^{+0.29}_{-0.20}$ &$[-2.52,1.75]$     & $0.60^{+0.28}_{-0.17}$ &$[-2.58,1.86]$   & $0.57^{+0.26}_{-0.16}$ &$[-2.59,1.90]$\\
                        &$2\widetilde{\lambda}_B$ & $0.28^{+0.15}_{-0.09}$ &$[-2.93,2.33]$     & $0.21^{+0.13}_{-0.07}$ &$[-2.93,2.38]$   & $0.20^{+0.11}_{-0.07}$ &$[-2.93,2.40]$\\
                        &BIK                      & $0.76$                 &                   & 0.63                   &                 & 0.59                   & \\ \hline

    $A_1$               &$\widetilde{\lambda}_B$  & $0.43^{+0.19}_{-0.12}$ &$[-1.38,0.49]$     & $0.37^{+0.16}_{-0.11}$ &$[-1.40,0.53]$   & $0.35^{+0.17}_{-0.10}$ &$[-1.43,0.54]$\\
                        &$2\widetilde{\lambda}_B$ & $0.17^{+0.09}_{-0.05}$ &$[-1.78,0.65]$     & $0.13^{+0.07}_{-0.04}$ &$[-1.75,0.67]$   & $0.12^{+0.07}_{-0.04}$ &$[-1.76,0.69]$\\
                        &BIK                      & $0.46$                 &                   & 0.38                   &                 & 0.36                   & \\ \hline

   $-\widetilde{A}_2$   &$\widetilde{\lambda}_B$  & $0.18^{+0.06}_{-0.05}$ &$[-3.47,3.64]$     & $0.14^{+0.06}_{-0.04}$ &$[-3.54,3.81]$   & $0.13^{+0.05}_{-0.04}$ &$[-3.57,3.88]$\\
                        &$2\widetilde{\lambda}_B$ & $0.07^{+0.04}_{-0.02}$ &$[-3.95,4.62]$     & $0.05^{+0.03}_{-0.01}$ &$[-3.95,4.68]$   & $0.05^{+0.03}_{-0.02}$ &$[-3.95,4.69]$\\
                        &BIK                      & $0.20$                 &                   & 0.16                   &                 & 0.14                   & \\ \hline

   $\widetilde{A}_3$    &$\widetilde{\lambda}_B$  & $1.07^{+0.52}_{-0.36}$ &$[-2.16,1.52]$     & $0.89^{+0.51}_{-0.35}$ &$[-2.18,1.59]$   & $0.86^{+0.50}_{-0.34}$ &$[-2.23,1.61]$ \\
                        &$2\widetilde{\lambda}_B$ & $0.28^{+0.26}_{-0.15}$ &$[-1.88,2.93]$     & $0.20^{+0.23}_{-0.14}$ &$[-1.71,4.03]$   & $0.19^{+0.22}_{-0.12}$ &$[-1.81,3.11]$ \\
                        &BIK                      & $1.17$                 &                   & 0.94                   &                 & 0.90                   & \\ \hline

   $T_1$                &$\widetilde{\lambda}_B$  & $0.54^{+0.22}_{-0.15}$ &$[-2.45,1.67]$     & $0.46^{+0.21}_{-0.14}$ &$[-2.51,1.78]$   & $0.44^{+0.20}_{-0.13}$ &$[-2.54,1.82]$\\
                        &$2\widetilde{\lambda}_B$ & $0.21^{+0.11}_{-0.07}$ &$[-2.86,2.22]$     & $0.16^{+0.09}_{-0.05}$ &$[-2.86,2.28]$   & $0.15^{+0.09}_{-0.05}$ &$[-2.87,2.30]$\\
                        &BIK                      & $0.57$                 &                   & 0.47                   &                 & 0.45                   & \\ \hline

   $T_2$                &$\widetilde{\lambda}_B$  & $0.54^{+0.23}_{-0.15}$ &$[-1.32,0.49]$     & $0.46^{+0.21}_{-0.14}$ &$[-1.34,0.52]$   & $0.44^{+0.20}_{-0.13}$ &$[-1.37,0.53]$\\
                        &$2\widetilde{\lambda}_B$ & $0.21^{+0.12}_{-0.07}$ &$[-1.71,0.61]$     & $0.16^{+0.10}_{-0.05}$ &$[-1.68,0.64]$   & $0.15^{+0.09}_{-0.05}$ &$[-1.70,0.65]$\\
                        &BIK                      & $0.58$                 &                   & 0.47                   &                 & 0.45                   & \\ \hline

   $\widetilde{T}_3$    &$\widetilde{\lambda}_B$  & $0.09^{+0.04}_{-0.03}$ &$[-1.86,1.11]$     & $0.07^{+0.03}_{-0.04}$ &$[-1.93,1.14]$   & $0.06^{+0.03}_{-0.03}$ &$[-1.95,1.20]$\\
                        &$2\widetilde{\lambda}_B$ & $0.09^{+0.03}_{-0.03}$ &$[-2.74,1.86]$     & $0.07^{+0.03}_{-0.02}$ &$[-2.80,2.02]$   & $0.06^{+0.03}_{-0.01}$ &$[-2.80,2.05]$\\
                        &BIK                      & $0.08$                 &                   & 0.06                   &                 & 0.06                   & \\ \hline

    $A_2$               &$\widetilde{\lambda}_B$  & $0.45^{+0.26}_{-0.18}$ &                   & $0.38^{+0.26}_{-0.18}$ &                 & $0.37^{+0.25}_{-0.17}$ &\\
                        &$2\widetilde{\lambda}_B$ & $0.11^{+0.13}_{-0.08}$ &                   & $0.08^{+0.11}_{-0.07}$ &                 & $0.07^{+0.11}_{-0.06}$ &\\ \hline

    $A_3$               &$\widetilde{\lambda}_B$  & $0.40^{+0.57}_{-0.37}$ &                   & $0.35^{+0.56}_{-0.39}$ &                 & $0.32^{+0.59}_{-0.37}$ &\\
                        &$2\widetilde{\lambda}_B$ & $0.25^{+0.27}_{-0.16}$ &                   & $0.21^{+0.24}_{-0.15}$ &                 & $0.20^{+0.25}_{-0.14}$ &\\ \hline

    $T_3$               &$\widetilde{\lambda}_B$  & $0.45^{+0.23}_{-0.15}$ &                   & $0.39^{+0.21}_{-0.14}$ &                 & $0.38^{+0.20}_{-0.13}$ & \\
                        &$2\widetilde{\lambda}_B$ & $0.12^{+0.12}_{-0.08}$ &                   & $0.09^{+0.10}_{-0.05}$ &                 & $0.09^{+0.09}_{-0.05}$ &\\ \hline
  \hline
\end{tabular}
\end{center}
\caption{ The values of the form-factors at zero  momentum transfer
and the parameters of the $B\to T$ form-factors, where the  BIK
denotes the $\phi_{+}(\omega)$ is taken as the BIK light-cone
distribution amplitude.}
\end{table}

\begin{table}
\begin{center}
\begin{tabular}{|cc|c|c|c|}\hline\hline
                                  &$\omega_0$               & $\widehat{F_{BK^*_2}}(0)$  & $\widehat{F_{Ba_2}}(0)$   & $\widehat{F_{Bf_2}}(0)$   \\ \hline
    $\widehat{V}$                 &$\widetilde{\lambda}_B$  & $1.00$                     & $1.00$                    & $1.00$ \\
                                  &$2\widetilde{\lambda}_B$ & $1.00$                     & $1.00$                    & $1.00$ \\
                                  &BIK                      & $1.00$                     & $1.00$                    & $1.00$   \\ \hline

    $\widehat{A_1}$               &$\widetilde{\lambda}_B$  & $0.61$                     & $0.62$                    & $0.61$ \\
                                  &$2\widetilde{\lambda}_B$ & $0.61$                     & $0.62$                    & $0.60$ \\
                                  &BIK                      & $0.61$                     & $0.60$                    & $0.61$  \\ \hline

   $\widehat{-\widetilde{A}_2}$   &$\widetilde{\lambda}_B$  & $0.25$                     & $0.23$                    & $0.23$ \\
                                  &$2\widetilde{\lambda}_B$ & $0.25$                     & $0.24$                    & $0.25$ \\
                                  &BIK                      & $0.26$                     & $0.25$                    & $0.24$  \\ \hline

   $\widehat{\widetilde{A}_3}$    &$\widetilde{\lambda}_B$  & $1.51$                     & $1.48$                    & $1.51$  \\
                                  &$2\widetilde{\lambda}_B$ & $1.00$                     & $0.95$                    & $0.95$  \\
                                  &BIK                      & $1.54$                     & $1.49$                    & $1.53$   \\ \hline

   $\widehat{T_1}$                &$\widetilde{\lambda}_B$  & $0.76$                     & $0.77$                    & $0.77$ \\
                                  &$2\widetilde{\lambda}_B$ & $0.75$                     & $0.76$                    & $0.75$ \\
                                  &BIK                      & $0.75$                     & $0.75$                    & $0.76$  \\ \hline

   $\widehat{T_2}$                &$\widetilde{\lambda}_B$  & $0.76$                     & $0.77$                    & $0.77$ \\
                                  &$2\widetilde{\lambda}_B$ & $0.75$                     & $0.76$                    & $0.75$ \\
                                  &BIK                      & $0.76$                     & $0.75$                    & $0.76$  \\ \hline

   $\widehat{\widetilde{T}_3}$    &$\widetilde{\lambda}_B$  & $0.13$                     & $0.12$                    & $0.11$ \\
                                  &$2\widetilde{\lambda}_B$ & $0.32$                     & $0.33$                    & $0.30$ \\
                                  &BIK                      & $0.11$                     & $0.10$                    & $0.10$  \\ \hline

    $\widehat{A_2}$               &$\widetilde{\lambda}_B$  & $0.63$                     & $0.63$                    & $0.65$ \\
                                  &$2\widetilde{\lambda}_B$ & $0.39$                     & $0.38$                    & $0.35$ \\ \hline

    $\widehat{A_3}$               &$\widetilde{\lambda}_B$  & $0.56$                     & $0.58$                    & $0.56$ \\
                                  &$2\widetilde{\lambda}_B$ & $0.89$                     & $1.00$                    & $1.00$ \\ \hline

    $\widehat{T_3}$               &$\widetilde{\lambda}_B$  & $0.63$                     & $0.65$                    & $0.67$  \\
                                  &$2\widetilde{\lambda}_B$ & $0.43$                     & $0.43$                    & $0.45$ \\ \hline
  \hline
\end{tabular}
\end{center}
\caption{ The central values of the ratios among the form-factors at
zero momentum transfer and the parameters of the form-factors, where
the BIK denotes the $\phi_{+}(\omega)$ is taken as the BIK
light-cone distribution amplitude,
$\widehat{F_{BK^*_2}}(0)=\frac{F_{BK^*_2}(0)}{V_{BK^*_2}(0)}$,
$\widehat{F_{Ba_2}}(0)=\frac{F_{Ba_2}(0)}{V_{Ba_2}(0)}$ and
$\widehat{F_{Bf_2}}(0)=\frac{F_{Bf_2}(0)}{V_{Bf_2}(0)}$.}
\end{table}

\begin{table}
\begin{center}
\begin{tabular}{|c|c|c|c|c|c|c|}\hline\hline
                      &Ref.\cite{ISGW2}  & Ref.\cite{CCH} & Ref.\cite{kcy}  & Refs.\cite{Charles,Ebert,Datta} & Ref.\cite{Wei}           & This Work\\ \hline
    $V(0)$            &0.38              & 0.29           & $0.16 \pm 0.04$ &$0.21\pm  0.03$                  & $0.21^{+0.06}_{-0.05}$   & $0.71^{+0.29}_{-0.20}$ $\left[0.28^{+0.15}_{-0.09}\right]$\\ \hline
  $A_1(0)$            &0.24              & 0.22           & $0.14 \pm 0.04$ &$0.14\pm  0.02$                  & $0.13^{+0.04}_{-0.03}$   & $0.43^{+0.19}_{-0.12}$ $\left[0.17^{+0.09}_{-0.05}\right]$\\ \hline
   $A_2(0)$           &0.22              & 0.21           & $0.05 \pm 0.04$ &$0.14\pm  0.02$                  & $0.08^{+0.03}_{-0.02}$   & $0.45^{+0.26}_{-0.18}$ $\left[0.11^{+0.13}_{-0.08}\right]$\\ \hline
  $A_0(0)$            &0.27              & 0.23           & $0.25 \pm 0.06$ &$0.15\pm  0.02$                  & $0.18^{+0.05}_{-0.04}$   & $0.40^{+0.57}_{-0.37}$ $\left[0.25^{+0.27}_{-0.16}\right]$\\ \hline
  $\widehat{A_1}(0)$  &0.63              & 0.76           & $0.88$          &$0.67$                           & $0.62$                   & $0.61$                 $\left[0.61\right]$\\ \hline
   $\widehat{A_2}(0)$ &0.58              & 0.72           & $0.31$          &$0.67$                           & $0.38$                   & $0.63$                 $\left[0.39\right]$\\ \hline
  $\widehat{A_0}(0)$  &0.71              & 0.79           & $1.56$          &$0.71$                           & $0.86$                   & $0.56$                 $\left[0.89\right]$\\ \hline
   \hline
\end{tabular}
\end{center}
\caption{ The values of the $B\to K^*_2(1430)$ form-factors at zero
momentum transfer from different theoretical approaches, the values
in the bracket  correspond to $\omega_0=2\widetilde{\lambda}_B$. We
take the central values of the ratios
$\widehat{A_1}(0)=\frac{A_1(0)}{V(0)}$,
$\widehat{A_2}(0)=\frac{A_2(0)}{V(0)}$ and
$\widehat{A_0}(0)=\frac{A_0(0)}{V(0)}$.}
\end{table}

\begin{table}
\begin{center}
\begin{tabular}{|c|c|c|c|c|c|c|}\hline\hline
                      &Ref.\cite{ISGW2}  & Ref.\cite{CCH} & Ref.\cite{kcy}    & Refs.\cite{Charles,Ebert,Datta}    & Ref.\cite{Wei}           & This Work\\ \hline
  $V(0)$              &0.32              & 0.28           & $0.18 \pm 0.02$   & $0.18\pm  0.03$                    & $0.18^{+0.05}_{-0.04}$   & $0.60^{+0.28}_{-0.17}$ $\left[0.21^{+0.13}_{-0.07}\right]$\\ \hline
  $A_1(0)$            &0.16              & 0.21           & $0.14 \pm 0.02$   & $0.13\pm  0.02$                    & $0.11^{+0.03}_{-0.03}$   & $0.37^{+0.16}_{-0.11}$ $\left[0.13^{+0.07}_{-0.04}\right]$\\ \hline
   $A_2(0)$           &0.14              & 0.19           & $0.09 \pm 0.02$   & $0.13\pm  0.02$                    & $0.06^{+0.02}_{-0.01}$   & $0.38^{+0.26}_{-0.18}$ $\left[0.08^{+0.11}_{-0.07}\right]$\\ \hline
  $A_0(0)$            &0.20              & 0.24           & $0.21 \pm 0.04$   & $0.14\pm  0.02$                    & $0.18^{+0.06}_{-0.04}$   & $0.35^{+0.56}_{-0.39}$ $\left[0.21^{+0.24}_{-0.15}\right]$\\ \hline
  $\widehat{A_1}(0)$  &0.50              & 0.75           & $0.78$            & $0.72$                             & $0.61$                   & $0.62$                 $\left[0.62\right]$\\ \hline
   $\widehat{A_2}(0)$ &0.44              & 0.68           & $0.50$            & $0.72$                             & $0.33$                   & $0.63$                 $\left[0.38\right]$\\ \hline
  $\widehat{A_0}(0)$  &0.63              & 0.86           & $1.17$            & $0.78$                             & $1.00$                   & $0.58$                 $\left[1.00\right]$\\ \hline
  \hline
\end{tabular}
\end{center}
\caption{ The values of the $B\to a_2(1320)$ form-factors at zero
momentum transfer from different theoretical approaches, the values
in the bracket  correspond to $\omega_0=2\widetilde{\lambda}_B$. We
take the central values of the ratios
$\widehat{A_1}(0)=\frac{A_1(0)}{V(0)}$,
$\widehat{A_2}(0)=\frac{A_2(0)}{V(0)}$ and
$\widehat{A_0}(0)=\frac{A_0(0)}{V(0)}$.}
\end{table}

\begin{table}
\begin{center}
\begin{tabular}{|c|c|c|c|c|c|c|}\hline\hline
                      &Ref.\cite{ISGW2}  & Ref.\cite{CCH}  & Ref.\cite{kcy}    & Refs.\cite{Charles,Ebert,Datta}  & Ref.\cite{Wei}          & This Work\\ \hline
  $V(0)$              &0.32              & 0.28            & $0.18 \pm 0.02$   & $0.18\pm  0.02$                  & $0.12^{+0.03}_{-0.03}$  & $0.57^{+0.26}_{-0.16}$ $\left[0.20^{+0.11}_{-0.07}\right]$\\ \hline
  $A_1(0)$            &0.16              & 0.21            & $0.14 \pm 0.02$   & $0.12\pm  0.02$                  & $0.08^{+0.02}_{-0.02}$  & $0.35^{+0.17}_{-0.10}$ $\left[0.12^{+0.07}_{-0.04}\right]$\\ \hline
   $A_2(0)$           &0.14              & 0.19            & $0.10\pm 0.02$    & $0.13\pm  0.02$                  & $0.04^{+0.01}_{-0.01}$  & $0.37^{+0.25}_{-0.17}$ $\left[0.07^{+0.11}_{-0.06}\right]$\\ \hline
  $A_0(0)$            &0.20              & 0.25            & $0.20 \pm 0.04$   & $0.13\pm  0.02$                  & $0.13^{+0.04}_{-0.03}$  & $0.32^{+0.59}_{-0.37}$ $\left[0.20^{+0.25}_{-0.14}\right]$\\ \hline
  $\widehat{A_1}(0)$  &0.50              & 0.75            & $0.78$            & $0.67$                           & $0.67$                  & $0.61$                 $\left[0.60\right]$\\ \hline
   $\widehat{A_2}(0)$ &0.44              & 0.68            & $0.56$            & $0.72$                           & $0.33$                  & $0.65$                 $\left[0.35\right]$\\ \hline
  $\widehat{A_0}(0)$  &0.63              & 0.89            & $1.11$            & $0.72$                           & $1.08$                  & $0.56$                 $\left[1.00\right]$\\ \hline
  \hline
\end{tabular}
\end{center}
\caption{ The values of the $B\to f_2(1270)$ form-factors at zero
momentum transfer from different theoretical approaches, the values
in the bracket  correspond to $\omega_0=2\widetilde{\lambda}_B$. We
take the central values of the ratios
$\widehat{A_1}(0)=\frac{A_1(0)}{V(0)}$,
$\widehat{A_2}(0)=\frac{A_2(0)}{V(0)}$ and
$\widehat{A_0}(0)=\frac{A_0(0)}{V(0)}$.}
\end{table}

In calculation, we observe that  the dominating contributions in the
sum rules for the form-factors come from the two-particle $B$-meson
light-cone distribution amplitudes, the contributions from the
three-particle $B$-meson light-cone distribution amplitudes are of
minor importance \footnote{The form-factor $\widetilde{T}_3(q^2)$ is
defined by $\widetilde{T}_3(q^2)=T_2(q^2)-T_3(q^2)\left(1-
q^2/\left(m_B^2-m_{K^*_2}^2\right)\right)$. In the sum rules for the
$T_2(q^2)$ and $\widetilde{T}_3(q^2)$, the dominating contributions
come from the two-particle and three-particle $B$-meson light-cone
distribution amplitudes, respectively. It is obvious that  the
form-factor $T_3(q^2)$ acquires  its value mainly from the
two-particle $B$-meson light-cone distribution amplitudes. }. This
is the prominent advantage of the $B$-meson light-cone QCD sum
rules. It is not un-expected that the dominating uncertainty comes
from the parameter $\omega_0$ (or $\lambda_B$), which determines the
line-shapes of the two-particle and three-particle light-cone
distribution amplitudes. We can take the value
$\omega_0=\lambda_B=(0.46\pm0.11)\,\rm{GeV}$ from the QCD sum rules
in Ref.\cite{Braun2004}, where the $B$-meson light-cone distribution
amplitude $\phi_+(\omega)$ is parameterized by the matrix element of
the bilocal operator at imaginary light-cone separation.

In Tables 3-5, we also present the values (and the ratios among the
central values) of the $B \to T$ form-factors from other theoretical
calculations, such as the improved version of the ISGW model
\cite{ISGW2}, the covariant light-front quark model
 \cite{CCH}, the light-cone sum rules   approach
\cite{kcy},  the large energy effective theory
\cite{Charles,Ebert,Datta} and the perturbative QCD approach
\cite{Wei}. From those tables, we can see that the central values of
the present predictions are at least (or almost) twice as large as
the existing estimations. If we take the $\phi_{+}(\omega)$ to be
the more elaborated model (the BIK model) suggested in
Ref.\cite{Braun2004}, the values of the $B\to T$ form-factors at
$q^2=0$ are even larger, see Table 1, where only the central values
are presented for simplicity. The two-particle $B$-meson
distribution amplitudes in the $k_T$ factorization formalism have
the same  Gaussian-like distribution amplitudes $\phi_{+}(\omega)$
and $\phi_{-}(\omega)$ besides additional factors describing the
$b$-parameter dependence \cite{Bdis2}. In this article, we do not
factorize out  the $k_T$ explicitly,  and cannot  take the
distribution amplitudes in $k_T$ factorization formalism.

The form-factors $T_1(q^2)$, $T_2(q^2)$ and $T_3(q^2)$ are related
to the radiative decays $B \to T\gamma$.
The branching ratio  of the radiative decay $B\to K_2^*(1430)\gamma$  can be written as
\begin{eqnarray}
  {\mathcal B}(B\to K_2^*\gamma)&=& \tau_{B}\frac{G_F^2 \alpha_{\rm em}m_B^5m_b^2|V_{tb}V_{ts}^*|^2}{256\pi^4
 m_{K_2^*}^2}\left(1-\frac{m_{K^*_2}^2}{m_B^2}\right)^5|C_7+A^{(1)}|^2|T_1^{BK^*_2}(0)|^2,
\end{eqnarray}
where the $V_{tb}$ and $V_{ts}$ are the CKM matrix elements, the $C_{7}$ is the Wilson coefficient for the operator $O_{7\gamma}$,
 the $A^{(1)}$ is the perturbative $\mathcal{O}(\alpha_s)$ corrections \cite{Ali2002}, the
$\tau_{B}$ is the $B$-meson  lifetime.
In Eq.(27), we take into account the factorizable contribution and  its perturbative $\mathcal{O}(\alpha_s)$ corrections
 in the standard model.
   From the experimental data
of the radiative decays $B^+ \to K_2^{*+}(1430)\gamma$ and $B^0 \to
K_2^{*0}(1430)\gamma$ \cite{BK2-1,BK2-2}, we can obtain the value
$T_1^{BK_2^*}(0)=0.16\pm0.02^{+0.00}_{-0.01}$, which relates to the
parameter  $\zeta_{\perp}^{BK_2^*}(0)=\frac{\vec{p}_{K_2^*}}{m_{K_2^*}}T_1^{BK_2^*}(0)=0.27\pm0.03^{+0.00}_{-0.01}$
in Refs.\cite{T-Difin-1,T-Difin-2}, the errors originate  from  the uncertainties of the experimental data and the $b$-quark
 pole mass, respectively. Compared  with the experimental value,  the present prediction
$T_1(0)=0.54^{+0.22}_{-0.15}$ seems too large, see Table 1.

There are two possible explanations  for the apparent discrepancy. One possibility  is that the non-factorizable contributions we have 
neglected in extracting the form-factor $T_1^{BK_2^*}(0)$ from the experimental data (for example, the contributions of the diagrams where the soft gluon is
emitted from the intermediate  $c\bar{c}$ loops and then absorbed by the $B$-meson) are large enough. If we take into account them consistently,
 the value extracted from the experimental data maybe compatible  with the present calculation.
 The non-factorizable contributions of the soft-gluon  emitted from the intermediate  $c\bar{c}$ loops  have been studied for
 the  $B\to K\gamma$ and $B\to K^*\gamma$ decays \cite{Khodjamirian2010},  while the corresponding  non-factorizable soft contributions of the
  $B \to K_2^*(1430)\gamma$  decay have not been calculated yet. The other possibility is that the non-factorizable contributions in the decay
  $B \to K_2^*(1430)\gamma$  are small enough to be neglected, the  form-factor $T_1^{BK_2^*}(0)$ extracted from the experimental data
  is precise enough, the apparent discrepancy is due to the shortcoming of the theory. In the following, we perform detailed discussions 
  about the second possibility.

In calculation, we observe that if we double the value of the
parameter $\omega_0$, i.e. taking
$\omega_0=\lambda_B=2\widetilde{\lambda}_B=(0.92\pm0.22)\,\rm{GeV}$
instead of
$\omega_0=\lambda_B=\widetilde{\lambda}_B=(0.46\pm0.11)\,\rm{GeV}$
(here we introduce a new parameter $\widetilde{\lambda}_B$ to avoid
confusion), we can obtain the value $T_1(0)=0.21^{+0.11}_{-0.07}$,
which is consistent with the theoretical estimations
$0.17^{+0.05}_{-0.04 }$, $0.19 \pm 0.04$ and $0.28$ from the
perturbative QCD \cite{Wei}, the light-cone QCD sum rules
\cite{Safir} and the covariant light-front quark model
\cite{ChengChua}, respectively, or the value
$0.16\pm0.02^{+0.00}_{-0.01}$ extracted from the experimental data
\cite{T-Difin-1,T-Difin-2}. The values (and the ratios among the
central values) of the $B\to T$ form-factors with $\omega_0=
2\widetilde{\lambda}_B$ are presented in Tables 1-5, from the
tables, we can see that the present predictions of the values of the
form-factors are compatible with other theoretical estimations,
while the ratios among the central values of the form-factors from
different theoretical approaches vary in a large range and no
definite conclusion can be made.

In the $B$-meson light-cone QCD sum rules for the $B\to\pi$, $K$,
$\rho,K^*$, $ D$, $D^*$, $a_1(1260)$, $K^*_0(1430)$, $a_0(1450)$
form-factors
\cite{Khodjamirian05,KhodjamirianB07,WangPLB,KhodjamirianEPJC,BScalar},
the main contributions come from the two-particle $B$-meson
light-cone distributions $\phi_\pm(\omega)$ as in the present case,
the form-factors $\propto \phi_\pm(\omega)$, the model light-cone
distribution
$\phi_+(\omega)=\frac{\omega}{\omega_0^2}\exp[-\frac{\omega}{\omega_0}]$
with the typical value $\omega_0=0.46\,\rm{GeV}$ can give
satisfactory results. In the present case, the form-factors $\propto
\omega\phi_+(\omega)$ due to the derivative in the interpolating
currents,  if the minor contributions are neglected, the predictions
based on  the typical value $\omega_0=0.46\,\rm{GeV}$ are too large.
There are two possible  reasons to account for the apparent
discrepancy: (1) the model light-cone distribution amplitudes
$\phi_\pm(\omega)$ are universal, but the $B$-meson light-cone QCD
sum rules are not applicable for the $B\to T$ form-factors, although
we can prove that the operator product expansion near the light-cone
$x^2\approx 0$ is feasible; (2) the $B$-meson light-cone QCD sum
rules are applicable for the $B\to T$ form-factors,  but the model
light-cone distribution amplitudes $\phi_\pm(\omega)$ are not the
ideal ones. A compromise  between these two reasons can be
suggested,  both the model light-cone distribution amplitudes
$\phi_\pm(\omega)$ and the $B$-meson light-cone QCD sum rules are
 robust, we can search for the ideal value of the parameter
$\omega_0$, as the line-shapes of the $B$-meson light-cone
distribution amplitudes have significant impacts on the values of
the form-factors.

The analytical  expressions of the form-factors are complicated, we
cannot obtain physical insight on   the $\omega_0$ dependence. We
can neglect the tiny  contributions from the three-particle
$B$-meson light-cone distribution amplitudes and other minor
contributions,  recast the expressions into simple form, and
 study the $\omega_0$ dependence. In
the following, we will present some typical examples for
illustration. The simplified expressions of the form-factors
$V_{B\to a_2}(0)$, $A_{B\to a_1}(0)$ and $F^+_{B\to \pi}(0)$ in the
transitions $B \to a_2(1320)$, $B\to a_1(1260)$ and $B\to \pi$
 can be written as
\begin{eqnarray}
V_{B\to a_2}(0) &=& \frac{f_B(m_B+m_{a_2})}{f_{a_2}
m_{a_2}^2}e^{\frac{m_{a_2}^2}{M^2}}\int_{0}^{\frac{s_0}{m_B}}d\omega
\omega\phi_+(\omega)e^{-\frac{\omega
m_B}{M^2}}\,, \nonumber\\
&=& \frac{\omega_0f_B(m_B+m_{a_2})}{f_{a_2}
m_{a_2}^2}e^{\frac{m_{a_2}^2}{M^2}}\int_{0}^{\frac{s_0}{\omega_{0}m_B}}dx
x^2e^{-\left(1+\frac{\omega_{0}
m_B}{M^2}\right)x}\,, \\
 A_{B\to a_1}(0)&=&\frac{f_B(m_B-m_{a_1})}{2f_{a_1} m_{a_1}} e^{\frac{m_{a_1}^2}{M^2} } \int_{0}^{\frac{s_0}{m_B}}d\omega
\phi_+(\omega)e^{-\frac{\omega
m_B}{M^2}}\,,\nonumber\\
&=&\frac{f_B(m_B-m_{a_1})}{2f_{a_1} m_{a_1}}
e^{\frac{m_{a_1}^2}{M^2} } \int_{0}^{\frac{s_0}{\omega_{0}m_B}}dx
xe^{-\left(1+\frac{\omega_{0}
m_B}{M^2}\right)x}\,, \\
F^+_{B\to \pi}(0)&=&\frac{f_B}{f_{\pi} } e^{\frac{m_{\pi}^2}{M^2} }
\int_{0}^{\frac{s_0}{m_B}}d\omega \phi_{-}(\omega)e^{-\frac{\omega
m_B}{M^2}}\,,\nonumber\\
&=&\frac{f_B}{f_{\pi} } e^{\frac{m_{\pi}^2}{M^2}
}\int_{0}^{\frac{s_0}{\omega_{0}m_B}}dx e^{-\left(1+\frac{\omega_{0}
m_B}{M^2}\right)x}\, ,
\end{eqnarray}
respectively, where the typical integral kernels are $\omega
\phi_{+}(\omega)$, $ \phi_{+}(\omega)$ and $ \phi_{-}(\omega)$,
respectively.  We can take the input parameters as
$m_{a_2}=1.31\,\rm{GeV}$, $f_{a_2}=0.107\,\rm{GeV}$,
$s^0_{a_2}=2.70\,\rm{GeV}$, $M_{a_2}^2=1.2\,\rm{GeV}^2$ in the
$a_2(1320)$ channel, $m_{a_1}=1.23\,\rm{GeV}$,
$f_{a_1}=0.238\,\rm{GeV}$, $s^0_{a_1}=2.55\,\rm{GeV}$,
$M_{a_1}^2=1.2\,\rm{GeV}^2$ in the $a_1(1260)$ channel and
$m_{\pi}=0.14\,\rm{GeV}$, $f_{\pi}=0.13\,\rm{GeV}$,
$s^0_{\pi}=0.70\,\rm{GeV}$, $M_{\pi}^2=1.0\,\rm{GeV}^2$ in the $\pi$
channel,  respectively
\cite{Khodjamirian05,KhodjamirianB07,WangPLB}, and plot the
numerical results in Fig.3. From the figure, we can see that the
values of the form-factors decrease monotonously with the  increase
of the $\omega_0$, in the region bellow the typical value
$\omega_0=0.46\,\rm{GeV}$ \cite{Braun2004}, the form-factors
decrease drastically, the curves are very steep, while in the region
above the typical value $\omega_0=0.46\,\rm{GeV}$, the form-factors
decrease more slowly, and the curves are flatter. If we take the
typical value $\omega_0=0.46\,\rm{GeV}$, the values of the
form-factors $ A_{B\to a_1}(0)$ and $F^+_{B\to \pi}(0)$ are
consistent with other theoretical estimations
\cite{Khodjamirian05,KhodjamirianB07,WangPLB}, while the value of
the form-factor $V_{B\to a_2}(0)$ is too large, on the other hand,
if we take the value
$\omega_0=2\widetilde{\lambda}_B=0.92\,\rm{GeV}$, the value of the
form-factor $V_{B\to a_2}(0)$ is consistent with other theoretical
estimations while the values of the form-factors $ A_{B\to a_1}(0)$
and $F^+_{B\to \pi}(0)$ are too small.

Irrespective of either of the two possible explanations for the apparent discrepancy between the
present theoretical calculation and the experimental extraction,
we should bear in mind that the values of the parameter $\omega_0$ from different theoretical
approaches differ  from each other greatly,
$\omega_0=(0.35\pm0.15)\,\rm{GeV}$ from the experiential value in
the QCD  factorization  \cite{BBNS}, $\omega_0=0.37\,\rm{GeV}$,
$0.46\pm0.11\,\rm{GeV}$, $0.6\,\rm{GeV}$ from different QCD sum
rules \cite{Grozin1997,Braun2004,BallKou}, $\omega_0=0.7\,\rm{GeV}$
from the Bakamjian-Thomas relativistic quark model \cite{BTmodel},
$\omega_0=(0.48\pm0.05)\,\rm{GeV}$ from the operator product
expansion \cite{Lee05}. The two form-factors in the decays $B^- \to
\gamma \ell^-\nu_{\ell}$ are proportional to $\frac{1}{\lambda_B}$
at the tree-level in the heavy quark limit, those processes are
directly related to the parameter $\lambda_B=\omega_0$  and would be
the most direct way of measuring it. In searching for the decays
$B^+ \to \gamma \ell^+\nu_{\ell}$, $\ell=e,\mu$, the Babar
collaboration has set the upper bounds $\omega_0>0.67\,\rm{GeV}$ or
$\omega_0>0.59\,\rm{GeV}$ \cite{Babar0704.1478,BTmodel},
$\omega_0>0.3\,\rm{GeV}$ \cite{Babar0907.1681}. It is difficult  to
choose the ideal value at the present time. All those theoretical calculations and experimental extractions concern
approximations in one or the other ways, and comprehensive theoretical analysis  are still needed.

\begin{figure}
 \centering
 \includegraphics[totalheight=4cm,width=5cm]{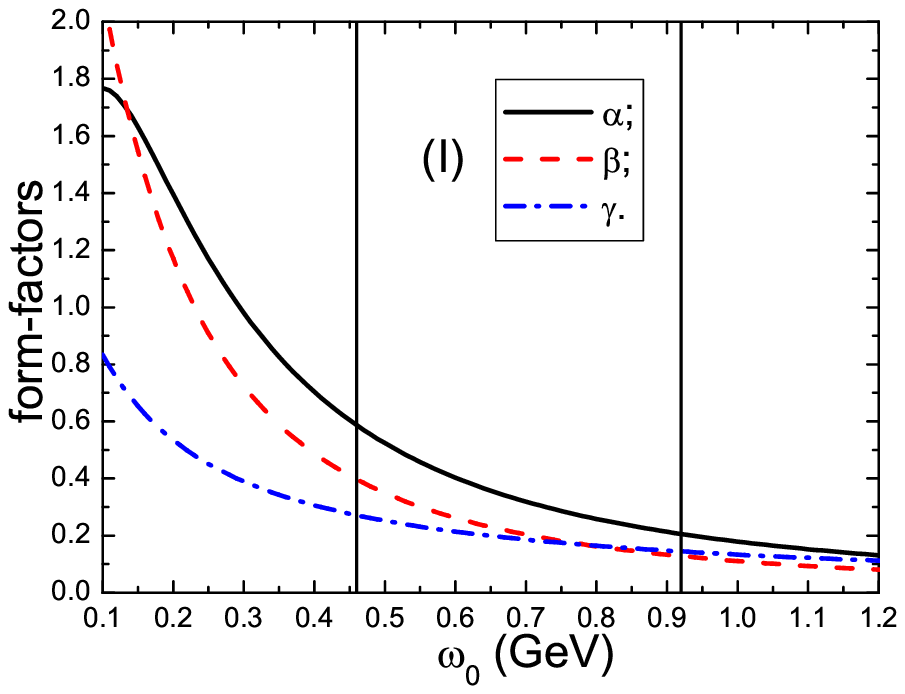}
 \includegraphics[totalheight=4cm,width=5cm]{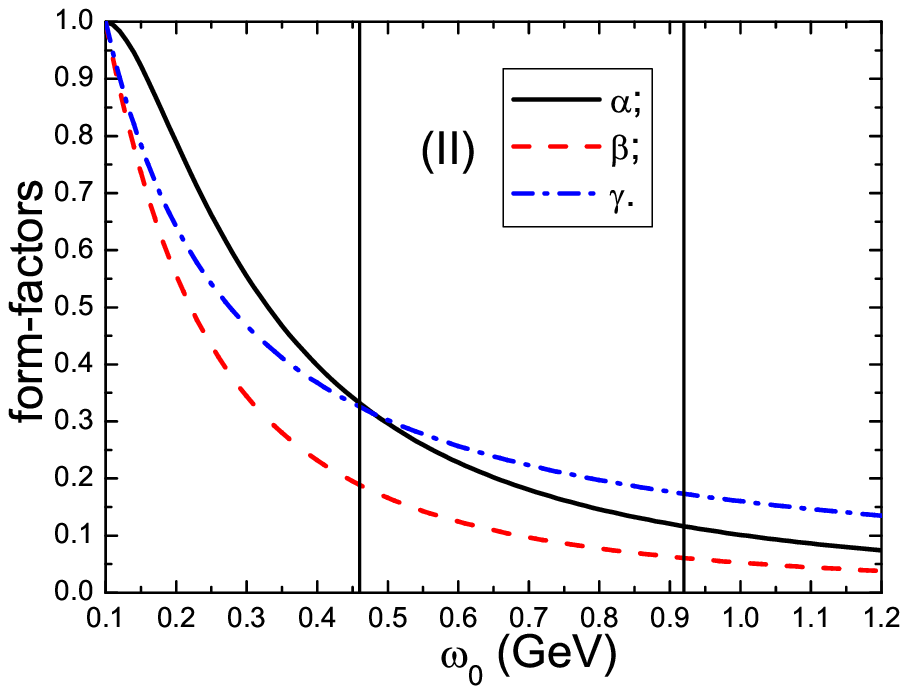}
    \caption{ The form-factors with variation of the parameter $\omega_0$ at zero momentum transfer,
the $\alpha$, $\beta$ and $\gamma$ denote the form-factors  $V_{B\to
a_2}(0)$, $A_{B\to a_1}(0)$ and $F^+_{B\to \pi}(0)$, respectively,
the two vertical lines correspond to the typical values
$\omega_0=0.46\,\rm{GeV}$ and $0.92\,\rm{GeV}$, respectively. In
(II), the form-factors are normalized to 1 at the value
$\omega_0=0.1\,\rm{GeV}$. }
\end{figure}

We can extract
 those form-factors from  the precise experimental data on the  radiative and
semi-leptonic decays $B\to T\gamma,\, T\ell\bar{\ell}$ at the KEK-B and LHCb in the future by including the
non-factorizable contributions, new physics effects, etc, in the theoretical analysis,  and
obtain severe constraints on the input parameter $\omega_0$ of the
$B$-meson  light-cone distribution amplitudes, although it is a hard
work. For example, the semi-leptonic decays $B\to a_2(1320)l \nu_l$
and $B\to f_2(1270)l \nu_l$ can be described by the effective
Hamiltonian at the lowest order approximation in the standard model,
 \begin{eqnarray}
 {\cal H}_{\rm eff}(b\to ul\bar \nu_l)=\frac{G_F}{\sqrt{2}}V_{ub}\bar
 u\gamma_{\mu}(1-\gamma_5)b \bar l\gamma^{\mu}(1-\gamma_5)\nu_l,
 \end{eqnarray}
where the $V_{ub}$ is the CKM matrix element and the $G_F$ is the
Fermi constant.  We can take the form-factors presented in Table 1
with $\omega_0=\lambda_B=2\widetilde{\lambda}_B$ as the input
parameters to study the partial (and total) decay widths $
d\Gamma/dq^2 $ (and $\int_{m_l^2}^{(m_{B}-m_{T})^2} dq^2
 (d\Gamma/dq^2)$),  and  the forward-backward (FB)
asymmetries $A_{FB}$ of the lepton,
 \begin{eqnarray}
 A_{FB}(q^2) &=& \frac{\int^{1}_{0} dz (d\Gamma/dq^2dz) -
\int^{0}_{-1} dz (d\Gamma/dq^2dz)}{\int^{1}_{0} dz (d\Gamma/dq^2dz)
+ \int^{0}_{-1} dz (d\Gamma/dq^2dz)}\, ,
 \end{eqnarray}
where $z=\cos\theta$ and the $\theta$ is the polar angle of the
lepton with respect to the moving direction of the tensor meson in
the lepton pair rest frame. Taking the other parameters from the
Review of Particle Physics \cite{PDG}, we can obtain the branching
ratios $1.6\times 10^{-4}$, $1.6\times 10^{-4}$, $0.6\times
10^{-4}$, $0.85\times 10^{-4}$, $0.85\times 10^{-4}$ and $0.34\times
10^{-4}$ for the semi-leptonic decays $\bar{B}^0\to a_2^+(1320)e
\bar{\nu}_e$, $\bar{B}^0\to a_2^+(1320)\mu \bar{\nu}_\mu$,
$\bar{B}^0\to a_2^+(1320)\tau \bar{\nu}_\tau$, $B^-\to f_2^0(1270)e
\bar{\nu}_e$, $B^-\to f_2^0(1270)\mu \bar{\nu}_\mu$ and $B^-\to
f_2^0(1270)\tau\bar{\nu}_\tau$, respectively, which are consistent
with the estimations from the perturbative QCD \cite{Wei}. The
forward-backward asymmetries $A_{FB}(q^2)$ are shown in Fig.4. In
this article, we show the central values explicitly.

\begin{figure}
 \centering
 \includegraphics[totalheight=4cm,width=5cm]{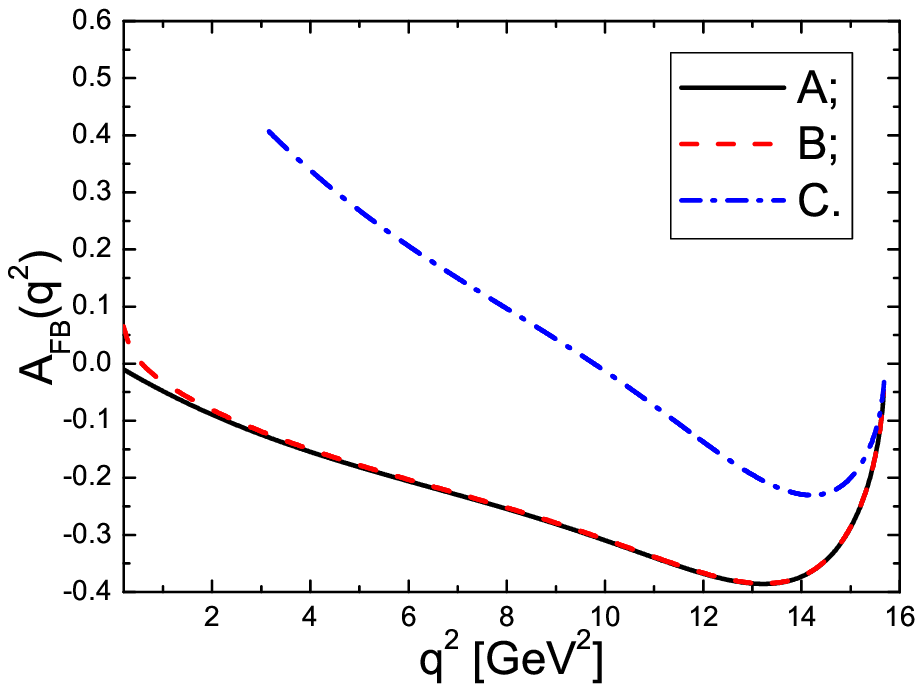}
 \includegraphics[totalheight=4cm,width=5cm]{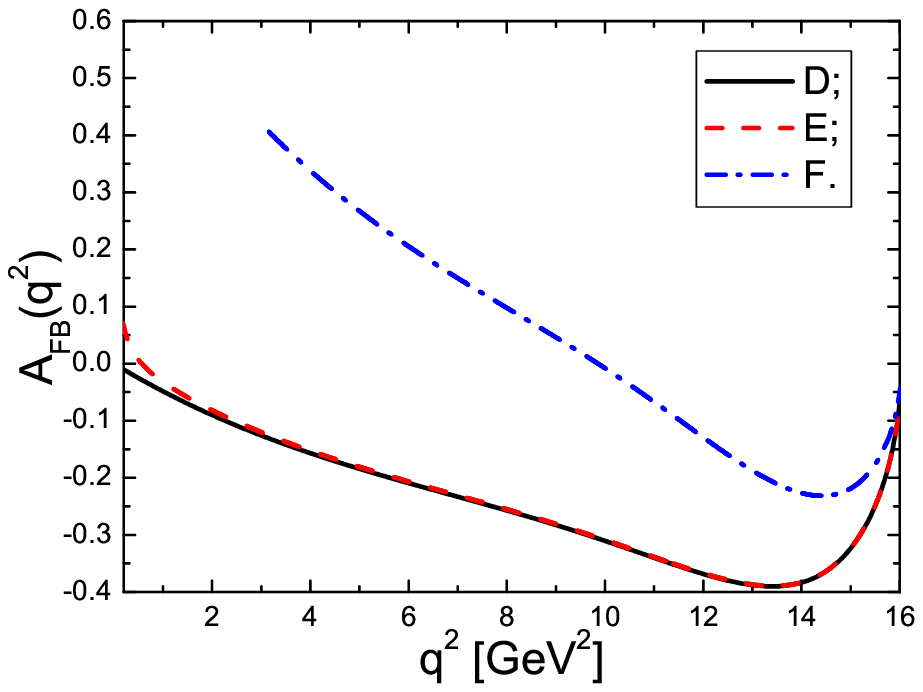}
    \caption{ The forward-backward asymmetries $A_{FB}(q^2)$ with the momentum transfer $q^2$,
     the $A$, $B$, $C$, $D$, $E$,  and $F$ correspond to the decays $\bar{B}^0\to
a_2^+(1320)e \bar{\nu}_e$, $\bar{B}^0\to a_2^+(1320)\mu
\bar{\nu}_\mu$, $\bar{B}^0\to a_2^+(1320)\tau \bar{\nu}_\tau$,
$B^-\to f_2^0(1270)e \bar{\nu}_e$, $B^-\to f_2^0(1270)\mu
\bar{\nu}_\mu$, and $B^-\to f_2^0(1270)\tau\bar{\nu}_\tau$,
respectively.    }
\end{figure}

\section{Conclusion}
In this article, we study the $B \to K^*_2(1430) ,  a_2(1320) ,
f_2(1270)$ form-factors $V(q^2)$, $A_1(q^2)$, $A_2(q^2)$, $
A_3(q^2)$, $T_1(q^2)$, $T_2(q^2)$ and $T_3(q^2)$
   with the $B$-meson light-cone QCD sum rules. In calculations, we
   observe that the dominating contributions come from the
   two-particle $B$-meson light-cone distribution amplitude $\phi_+(\omega)$, its line-shapes
   have significant impacts on the
values of the form-factors, we can search for the ideal values of
the parameter $\omega_0$. In the $B$-meson light-cone sum rules for
the $B\to P, V, A, S $ form-factors, the dominating contributions
$\propto \phi_\pm(\omega)$, while in the $B$-meson light-cone sum
rules for the $B\to T$ form-factors, the dominating contributions
$\propto\omega\phi_+(\omega)$. If we
   take the value $\omega_0=\lambda_B=\widetilde{\lambda}_B$ as in the $B$-meson light-cone QCD sum rules for the
   $B\to P, V, A, S $ form-factors, the central
values of the present predictions are at least (or almost) twice as
large as the existing theoretical estimations, and the $T_1(0)$
deviates  greatly from the value extracted from the radiative decays
$B^+ \to K_2^{*+}(1430)\gamma$ and $B^0 \to K_2^{*0}(1430)\gamma$, the non-factorizable contributions are neglected
in the extraction.
 On the other hand, if we take the value
$\omega_0=\lambda_B=2\widetilde{\lambda}_B$, the present predictions
are compatible with other theoretical estimations.
  In calculations, we observe that the main uncertainty comes from the parameter
  $\omega_0$ (or $\lambda_B$),
which determines  the line-shapes of  the two-particle and
three-particle $B$-meson light-cone distribution amplitudes, it is
of great importance to refine this parameter. We can extract the
values of those form-factors from the experimental data on the
radiative and semi-leptonic decays at the KEK-B and the LHCb in the
future, and obtain severe constraints on  the  parameter $\omega_0$.

\section*{Acknowledgments}
This  work is supported by National Natural Science Foundation of
China, Grant Number 11075053,  and the Fundamental Research Funds for the Central Universities.

\end{document}